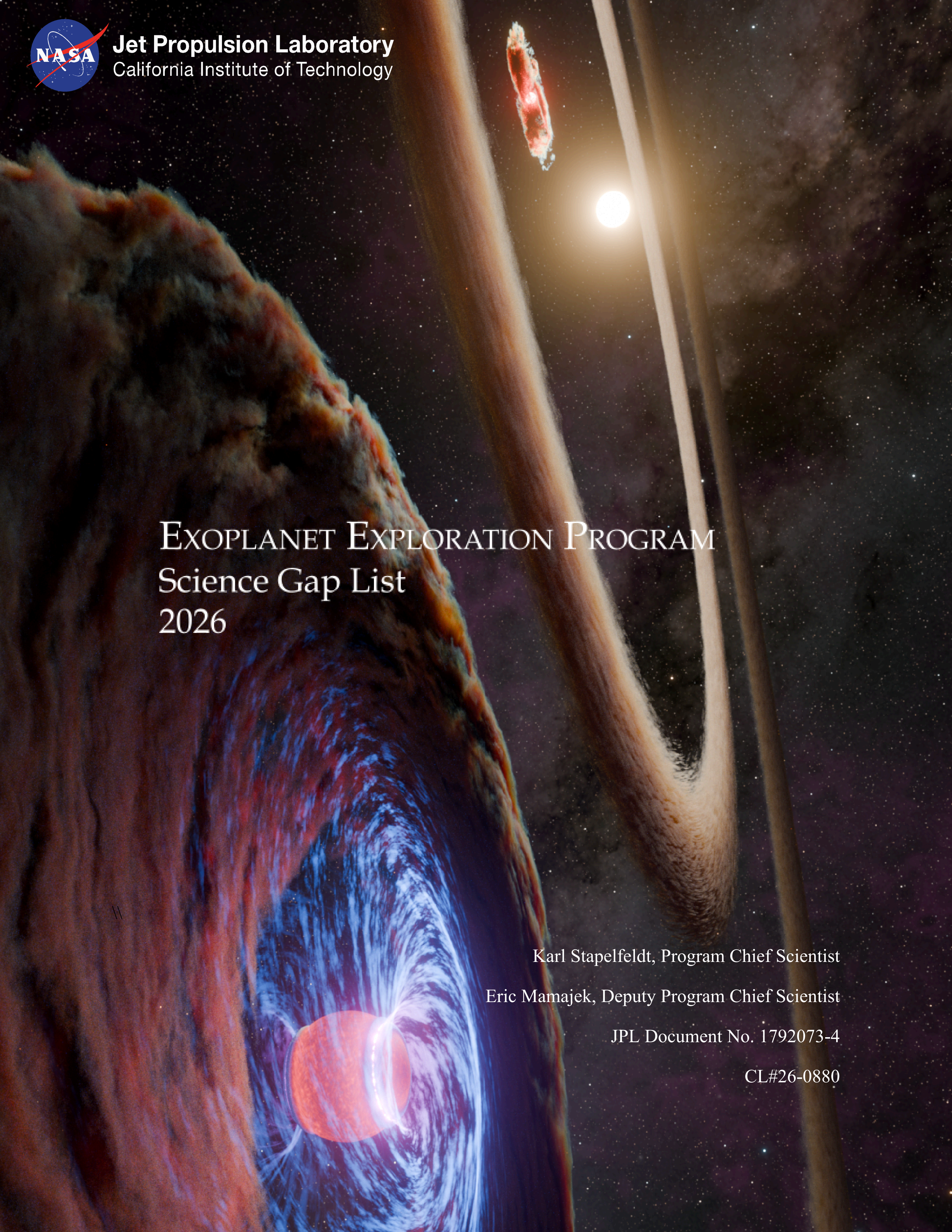

Jet Propulsion Laboratory
California Institute of Technology

# Exoplanet Exploration Program
## Science Gap List
## 2026

Karl Stapelfeldt, Program Chief Scientist

Eric Mamajek, Deputy Program Chief Scientist

Cover Art Credit: NASA/JPL-Caltech/Robert Hurt. Artist's concept depicting the young gas giant protoplanet WISPIT 2b accreting matter from its circumplanetary disk in the foreground, with the star WISPIT 2, two of its multiple circumstellar dust rings, and the inner gas giant protoplanet WISPIT 2c in the background. The WISPIT 2 system lies 433 light years away in the constellation Aquila.

The artwork accompanied a pair of papers describing the 2025 discoveries of the circumstellar disk rings and giant protoplanets orbiting the young star WISPIT 2 (TYC 5709-354-1) from the Wide Separation Planets In Time (WISPIT) survey. The journal article "*WIde Separation Planets In Time (WISPIT): A Gap-clearing Planet in a Multi-ringed Disk around the Young Solar-type Star WISPIT 2*" by R. van Capelleveen et al. (2025, Astrophysical Journal Letters, Vol. 990, L8) presented the discovery and analysis of the resolved, multi-ringed circumstellar disk and the protoplanet candidate WISPIT 2b, imaged using observations from the Very Large Telescope (VLT) Spectro-Polarimetric High-contrast Exoplanet Research (SPHERE) instrument at the European Southen Observatory (ESO). Evidence that the giant protoplanet was accreting from a circumplanetary disk, and preliminary evidence for another, closer-in giant protoplanet, was presented in the article "*Wide Separation Planets in Time (WISPIT): Discovery of a Gap Hα Protoplanet WISPIT 2b with MagAO-X*" by L. Close et al. (2025, Astrophysical Journal Letters, Vol. 990, L9) using observations with the MagAO-X (Magellan eXtreme Adaptive Optics) system on the Magellan Clay 6.5-m telescope at Las Campanas Observatory in Chile, and the LMIRCam infrared camera and coronagraph instrument on the Large Binocular Telescope Interferometer (LBTI) at Large Binocular Telescope Observatory on Mt. Graham in Arizona. The inner candidate protoplanet, now dubbed WISPIT 2c, was subsequently confirmed by Lawlor et al. (2026, arXiv:2603.22085) and estimated to be nearly twice as massive (~8-12 Jupiter masses) as planet b (~5 Jupiter masses). The two giant protoplanets b and c orbit their star WISPIT 2 at orbital separations of approximately 57 au and 14 au, respectively. The ~5 million-year-old WISPIT-2 system appears to be a scaled-up version of what theoretical simulations suggest our solar system with gas giants Jupiter and Saturn may have looked like in their late stages of formation.

An early clue to the existence of this baby exoplanetary system, unrecognized at the time, was the detection of excess infrared light from WISPIT 2's circumstellar dust by the IRAS (InfraRed Astronomical Satellite) space telescope in the early 1980s. IRAS launched in 1983 and was the first space telescope to map the sky in infrared light. The infrared source IRAS 19205-0746, corresponding to WISPIT 2, was first catalogued in the IRAS Point Source Catalog in 1988, but the nature of the excess infrared emission had to wait for future generations of large telescopes and instrumentation. IRAS was a joint project of NASA, the Netherlands Agency for Aerospace Programmes and the United Kingdom's Science and Engineering Research Council.

This research was carried out at the Jet Propulsion Laboratory, California Institute of Technology, under a contract with the National Aeronautics and Space Administration.

This document has been cleared for public release (CL#26-0880).

**Approved by:**

E-SIGNED by Dawn Gelino on 2026-03-27 22:45:07 GMT | 2026-03-27 22:45:07

Dr. Dawn Gelino
Program Manager
Exoplanet Exploration Program Office
NASA/Jet Propulsion Laboratory
California Institute of Technology

Date

E-SIGNED by David Ardila on 2026-03-27 23:08:44 GMT | 2026-03-27 23:08:44

Dr. David Ardila
Deputy Program Manager
Exoplanet Exploration Program Office
NASA/Jet Propulsion Laboratory
California Institute of Technology

Date

E-SIGNED by Hannah Jang-Condell on 2026-03-27 22:24:08 GMT | 2026-03-27 22:24:08

Dr. Hannah Jang-Condell
Program Scientist
Exoplanet Exploration Program
Science Mission Directorate
NASA Headquarters

Date

E-SIGNED by Joshua Pepper on 2026-03-31 01:09:46 GMT | 2026-03-31 01:09:46

Dr. Joshua Pepper
Deputy Program Scientist
Exoplanet Exploration Program
Science Mission Directorate
NASA Headquarters

Date

**Created by:**

E-SIGNED by Karl Stapelfeldt
on 2026-03-30 09:04:15 GMT

2026-03-30 09:04:15

Dr. Karl Stapelfeldt
Program Chief Scientist
Exoplanet Exploration Program Office
NASA/Jet Propulsion Laboratory
California Institute of Technology

Date

E-SIGNED by Eric Mamajek
on 2026-03-30 22:35:30 GMT

2026-03-30 22:35:30

Dr. Eric Mamajek
Deputy Program Chief Scientist
Exoplanet Exploration Program Office
NASA/Jet Propulsion Laboratory
California Institute of Technology

Date

# Table of Contents

# 1. Introduction to the 2026 Exoplanet Exploration Program (ExEP) Science Gap List

The Exoplanet Exploration Program (ExEP) is chartered by the Astrophysics Division (APD) of NASA's Science Mission Directorate (SMD) to carry out science, research, and technology tasks that advance NASA's science goals for exoplanets, which are:

- *Discover planets around other stars,*
- *Characterize their properties, and*
- *Identify candidates that could harbor life*

ExEP serves NASA and the community by acting as a focal point for exoplanet science and technology, managing research and technology initiatives, facilitating access to scientific data, and integrating the results of previous and current missions into a cohesive strategy to enable future discoveries. ExEP serves the critical function of developing the concepts and technologies for exoplanet missions, in addition to facilitating science investigations derived from those missions. ExEP manages development of mission concepts, including key technologies, as directed by NASA HQ, from their early conceptual phases into pre-Phase A.

The ExEP *Science Gap List* (SGL) tabulates program "science gaps", which are defined as either:

- *The difference between knowledge needed to define requirements for specified future NASA exoplanet missions and the current state of the art, or*

- *Knowledge which is needed to enhance the exoplanet science return of current and upcoming NASA exoplanet missions.*

Making the gap list public signals to the broader community where focused activities and investigations are needed over the next 3-5 years to advance the goals of NASA's Exoplanet Exploration Program. All ExEP approaches, activities, and decisions are guided by science priorities, and those priorities are presented and summarized in the ExEP Science Gap List[1].

The central community report relevant to the NASA ExEP is the *Pathways to Discovery in Astronomy and Astrophysics for the 2020s* (*Astro2020*) Decadal survey report from the National Academies of Sciences, Engineering, and Medicine released in November 2021. *Astro2020* included input from two other recent National Academies reports: the National Academies'

[1] The ExEP Science Gap List is one of three ExEP science plan documents maintained before Astro2020, the other two being the ExEP Science Plan Appendix and ExEP Science Development Plan. For descriptions of these documents, see the introduction to the 2023 edition of the ExEP Science Gap List. Previous versions of the ExEP Science Gap List and the two other documents (all obsolete) are available at
https://science.nasa.gov/astrophysics/programs/exep/resources/documents/
Updates to the Science Plan Appendix and Science Development Plan are under consideration.

*Exoplanet Science Strategy* (ESS) and *An Astrobiology Strategy for the Search for Life in the Universe*, both released in late 2018. *Pathways to Discovery in Astronomy and Astrophysics for the 2020s* identifies the most compelling science goals and presents an ambitious program of ground- and space-based activities to address them through investments in the 2020s and beyond. *Astro2020* identifies three major science themes for the next decade, the first of which (*Worlds and Suns in Context*) calls for investigations of Earth-like exoplanets. Two other themes focus on the most energetic processes in the universe and the evolution of galaxies.

The *Exoplanet Science Strategy* report provided a broad-based community assessment of the state of the field of exoplanet science and recommendations for future investments. NASA HQ's major response to the ESS report was to charter the "Extreme Precision Radial Velocity Working Group" (EPRV-WG), which developed and presented a blueprint for a strategic EPRV initiative to NASA and NSF in March 2020, and produced a final report in summer 2021[2].

The 2018 Exoplanet Science Strategy report provided seven recommendations, thirty-five findings, and two "*overarching goals in exoplanet science*":

- *to understand the formation and evolution of planetary systems as products of the process of star formation, and characterize and explain the diversity of planetary system architectures, planetary compositions, and planetary environments produced by these processes, and*

- *to learn enough about the properties of exoplanets to identify potentially habitable environments and their frequency and connect these environments to the planetary systems in which they reside. Furthermore, scientists need to distinguish between the signatures of life and those of nonbiological processes, and search for signatures of life on worlds orbiting other stars*

The recommendations from Astro2020, the 2018 *Exoplanet Science Strategy* and the 2018 *An Astrobiology Strategy for the Search for Life in the Universe* reports are all factored into the *2026 ExEP Science Gap List*. The "*highest priority for space frontier missions*" is a future large near-infrared/optical/ultraviolet space telescope optimized for observing habitable exoplanets and general astrophysics, nominally with diameter ~6 meters and capable of high-contrast ($\sim10^{-10}$) imaging and spectroscopy, and which is now referred to as the *Habitable Worlds Observatory (HWO)*[3].

The *Astro2020* Decadal Survey science theme *Worlds and Suns in Context* had two science panel discovery areas ("*The Search for Life on Exoplanets*" and "*Detecting and Characterizing Forming Planets*") and several questions from the Decadal Science Panels (Table 2.1 of Astro2020) which map to gaps in the ExEP Science Gap List:

[2] Extreme Precision Radial Velocity Working Group (EPRV WG) Final Report: https://arxiv.org/abs/2107.14291

[3] https://science.nasa.gov/astrophysics/programs/habitable-worlds-observatory/

- E-Q1: *What is the range of planetary system architectures, and is the configuration of the solar system common*?
- E-Q2: *What are the properties of individual planets, and which processes lead to planetary diversity*?
- E-Q3: *How do habitable environments arise and evolve within the context of their planetary systems*?
- E-Q4: *How can signs of life be identified and interpreted in the context of their planetary environments*?
- F-Q4: *Is planet formation fast or slow*?
- G-Q3: *What would stars look like if we view them like we do the Sun*?

The SGL notes the linkage between each ExEP Science Gap and these *Astro2020* science questions.

In April 2022 the National Academies released a second Decadal Survey *Origins, Worlds, and Life: A Decadal Strategy for Planetary Science and Astrobiology 2023-2032*. While it makes no recommendations for exoplanet missions or projects, its Chapter 12 provides an extensive list of exoplanet science questions and identifies areas for strategic research, anchored to the overall question "*What does our planetary system and its circumplanetary systems of satellites and rings reveal about other planetary systems, and what can disks and exoplanets orbiting other stars teach us about the solar system?*" We have noted connections between ExEP Science Gaps and these questions where appropriate.

The ExEP science gaps do *not* appear in a particular order, and by being recognized on this list they are deemed important. Currently the gap list is used when evaluating possible new Program activities: if a proposed activity could close a gap, it would be considered for greater priority for Program resources. The ExEP Science Gap List is *not* meant to provide strategic community guidance on par with a National Academies report (e.g., Decadal Surveys, Exoplanet Science Strategy, etc.), but to provide program-level tactical guidance for program management within the ever-shifting landscape of NASA missions and mission studies. Funding sources outside NASA ExEP are free to make their own judgments as to whether or not to align the work they support with NASA's Exoplanet Exploration goals. Science gaps specific to individual missions in phases A-E are relegated to those missions and are not tracked in the ExEP SGL. However, science gaps that facilitate science investigations derived from those missions, or support phase A or pre-phase A studies of new missions, may appear in the SGL.

Following usage in the two NASA Precursor Science Workshops ("Precursors to Pathways: Science Enabling NASA Astrophysics Future Great Observatories") held in April and October 2022, we adopt the following definitions to clarify the nature of science investigations, their relevance to NASA missions, and their urgency:

- **Precursor Science:** Informs mission design, architecture, and trade studies.
    - *When needed*: Generally, before a mission finalizes its architecture. For *HWO*, this is the present.

- *Example*: Astro 2010 Decadal Survey pointed out "*measurement of exozodiacal light levels*" that would "*determine the size and complexity*" of a mission capable of imaging rocky planets orbiting within the habitable zones of nearby stars.

- **Preparatory Science:** Informs early operations or interpretation of mission data.
  - *When needed*: By, or just after, launch.
  - *Example*: Completion of the Two Micron All-Sky Survey (2MASS) survey provided critical images and an all-sky catalog of near-infrared sources to support target definition and mission operations for the *Spitzer Space Telescope* launched in 2003. The 2MASS survey did not need to be completed while *Spitzer* was being formulated, so it was not "precursor science," but "preparatory science."

- **Follow-up Science:** Investigations that follow up on discoveries or other science from the mission.
  - *When needed*: After launch, but coordination and planning may be required prior to launch.
  - *Example*: Spectroscopic characterization and precision radial velocity measurements of stars hosting newly identified exoplanet candidates discovered with the *Kepler* and *TESS* missions, in order to measure the properties of the transiting planet (especially its mass).

Section 3 provides a list of ExoPAG Science Analysis Groups (SAGs) whose analyses on exoplanet-related topics were provided to NASA ExEP and the Astrophysics Division. SAG final reports are cited where relevant to the gap topic.

Section 4 provides a glossary of the acronyms commonly used by NASA ExEP which may be encountered among the SGL descriptions or other ExEP documents.

Section 5 discusses some adopted definitions for a few exoplanet terms.

The authors would like to acknowledge community members and especially members of the ExoPAG Executive Committee for their valuable comments & suggestions on the many versions of this gap list over the years. The value of this product has been greatly increased by their efforts.

# 2. The 2026 Exoplanet Exploration Program (ExEP) Science Gap List

## 2.1. SCI-01: Spectroscopic observations of the atmospheres of small exoplanets

- Connects to Astro2020 Decadal Science Panel Questions E-Q1, E-Q2, E-Q3, E-Q4
- Connects to PSA2022 Decadal Science Questions 12.3, 12.6, 12.10, 12.11
- *Related gaps:* limits to precision on extracting spectra (gap SCI-03), value of precursor surveys to find direct imaging targets (gap SCI-10).

**Gap Summary:**
The study of planetary atmospheres advances our knowledge of planetary formation and evolution. The Large Near-IR/Optical/UV space telescope recommended by Astro2020 (*Habitable Worlds Observatory, HWO*) is tasked with imaging and spectrally characterizing a robust sample of ~25 potentially temperate rocky exoplanets in the UV/visible/near-IR. To date there are a limited number of spectroscopic detections of atmospheres for exoplanets smaller than Neptune, even though they dominate the known exoplanet population. While some spectral constraints have been obtained for sub-Neptunes and super-Earths, detection of spectral features for temperate Earth-sized exoplanets has been beyond current capabilities. Initial *JWST* results indicate that small exoplanets are difficult targets for transit spectroscopy due to stellar contamination (gap SCI-15) or atmospheric erosion by the radiation environments of low-mass stars. To remotely assess the frequency of habitable planets and life, new observations and facilities must be developed to characterize the atmospheres of small exoplanets.

Spectroscopic observations of small exoplanets by *JWST* may provide precursor science results that influence the requirements for *HWO* instrumentation.

**Capability Needed:**
- Spectroscopy of small exoplanets across a diverse range of planet sizes, temperatures, atmospheric compositions, stellar types, and radiation environments, e.g., transit (transmission and emission) spectroscopy of small planets transiting cool dwarf stars, and high-contrast spectroscopy of small exoplanets orbiting solar-type (FGK-type) stars.
- Temperate examples are of particular interest for searching for biosignatures.
- Targets are needed that provide the most photons (i.e., orbiting nearby, brightest stars for their class).

**Capability Today:**
- *HST* and ground-based transit spectra have provided the constraints for some sub-Neptune sized planets but have marginal sensitivity only sufficient to detect spectral features in cloud-free H-dominated atmospheres.
- Kreidberg & Stevenson (2025, PNAS, 1222, 39) review early *JWST* results on transmission and emission spectroscopy of transiting rocky planets. Despite some tentative detections, and instances of stellar contamination, "*no atmospheres have been* *convincingly detected*," however many plausible atmospheric compositions and surface pressures are below the noise levels of the observations.

- To date *TESS* has identified ~315 small ($R < 3\ R_{Earth}$), mostly hot exoplanets suitable for spectroscopic followup.
- Hord et al. (2024, AJ, 167, 233) present an analysis of the *TESS* exoplanets discovered through ~2024 (both confirmed and candidates) identifying "best-in-class" planets most amenable to transmission spectroscopy and emission spectroscopy observations with *JWST*, ranked in bins by planet size and equilibrium temperature. Approximately half of the TESS candidates require confirmation.
- So far there are no imaging detections of small exoplanets.

**Mitigations in Progress:**

- Roughly sixty small exoplanets identified by transit surveys have been observed or approved for spectroscopic observations by *JWST* through Cycle 3. Results for these are now being reported in the literature.
- The *JWST* Director's Discretionary Time program "Rocky Worlds" and GO program "Hot Rocks Survey" are conducting time series emission photometry surveys at 15 µm to detect thermal emission from M dwarf rocky planets in order to detect atmospheres (Redfield et al., 2024; arXiv:2404.02932; August et al., 2025, A&A, 695, A171).
- NExScI tracks[4] and enables searches for published exoplanetary atmosphere spectra.
- High dispersion spectroscopy coupled with extreme AO coronagraphy is being developed at ESO and Keck and may provide some detections of hot planets smaller than Neptune, if their velocity amplitudes are large enough to distinguish their spectral lines from stellar and telluric features.
- Snellan (2025, ARA&A, 63, 83) provides a modern review of future instrumentation projects in development for the next generation of extremely large telescopes (e.g., ELT, GMT, TMT) with high resolution spectroscopy exoplanet science cases, and the prospects for detecting the atmospheres of nearby small exoplanets.
- The *Pandora* smallsat launched in February 2026 and will observe about 20 exoplanets[5] with simultaneous transit spectroscopy and optical photometry. Half a dozen of the *Pandora* targets are sub-Neptunes or super-Earths.

*Cross-Divisional Synergy:*

- The *OSIRIS-APEX* mission obtained Earth spectra as a function of phase angle during its gravity-assist flyby of Earth in September 2025. Analysis of these data should help set expectations for the brightness of habitable zone rocky planets when they are imaged by the *HWO* in the future. Additional Earth spectra to explore seasonal variability are intended during the upcoming 2027 and 2029 gravity-assist flybys.

[4] https://exoplanetarchive.ipac.caltech.edu/cgi-bin/atmospheres/nph-firefly?atmospheres
[5] *Pandora* target list at ExoFOP: https://exofop.ipac.caltech.edu/tess/view_targets_pandora.php

## 2.2. SCI-02: Modeling exoplanet atmospheres

- Connects to Astro2020 Decadal Science Panel Questions E-Q1, E-Q2, E-Q3, E-Q4
- Connects to PSA2022 Decadal Science Questions 12.6, 12.10
- *Related gaps: SCI-13 for laboratory measurements of the physical and chemical properties of molecules and aerosols, SCI-14 for exoplanet interior structure and material properties, SCI-16 for advancing biosignature research to improve interpretation of spectra of potentially habitable worlds.*

**Gap Summary:**

Spectral modeling is essential for inferring the properties of exoplanet atmospheres, identifying their most crucial diagnostics, and defining the design goals for future telescopes and instruments. There are many complexities to model approaches, initial conditions, chemistry, energetics, and the evolution of atmospheres such that many potentially relevant scenarios remain unexplored. Improving the fidelity of exoplanet atmosphere models and synthetic spectra can enhance the science return of *JWST* and *HST* observations (follow-up science), enhance the science return of *Roman* Coronagraph Instrument and *Ariel* (preparatory science), and refine understanding of the needed spectral resolution, S/N, and wavelength coverage that could influence the requirements for *HWO* to spectrally characterize exoplanets (precursor science).

**Capability Needed:**

- Ability to model the physical and chemical structure of exoplanet atmospheres and their emergent spectra across the range of planet masses, sizes, and stellar host types.
- Treat the effects of the total atmospheric pressure; chemical composition; presence of condensates, clouds & hazes; observer phase angle; and the radiative and energetic particle fluxes incident from the host star.
- Understand how the exchange of matter and energy between exospheres, lithospheres, hydrospheres, and potentially biospheres, affects the observed properties of the atmosphere today and over the planet's history.
- Challenges include determining composition of aerosols, understanding chemistry (e.g., relevant reactions, photochemistry, mixing), radiative transfer modeling (including scattering prescriptions), and 3D atmosphere dynamics (e.g., general circulation models).
- Modeling of the effects of greenhouse gases for assessment of surface temperatures and stability of surface water (habitability). Exploration of all the above parameters over a wide range of plausible exoplanetary atmospheres that might be discovered.

**Capability Today:**

- Thermophysical, radiative transfer, and photochemical models of planetary atmospheres in the solar system.
- Modeling of gas giant atmospheres accounting for varying formation mechanisms, protoplanetary disk chemistry, and migration.
- 3D circulation models of hot giant planets, modeling the impact of non-uniform cloud cover, modeling atmospheric chemistry and escape due to stellar XUV emission and predicted spectral observations (e.g., *HST, JWST*, future missions, etc.).
- Discrepancies have emerged between general circulation models that need to be reconciled (Fauchez et al., 2021, PSJ, 2, 106).

- Abiotic origins for molecular oxygen have been explored (e.g., June 2018 special issue of Astrobiology; Krissansen-Totten et al., 2022, AGU Advances, 2, 294).
- Krissansen-Totton et al. (2025; arXiv:2507.14771) present an analysis of coronagraph requirements for life detection via reflected light spectroscopy with *HWO*, proposing wavelength ranges and resolutions in the NUV, visible, and near-IR, respectively.

**Mitigations in Progress:**

- Ongoing research by the community, with support from XRP, *JWST*, and *HST* funding sources.
- The Exoplanet Modeling and Analysis Center (EMAC)[6] has been established at NASA/GSFC to serve as a catalog & repository for exoplanet atmosphere models.
- An Exoplanet Opacity Database[7] has been set up as a resource for the modeling community.

*Cross-Divisional Synergies*:

The Earth and solar system planets provide a prime opportunity for model validation. NASA ROSES XRP supports investigations exploring the remotely observable chemical and physical processes in exoplanet atmospheres, including theory.

[6] https://emac.gsfc.nasa.gov/
[7] https://science.data.nasa.gov/opacities/

## 2.3. SCI-03: Spectral signature retrieval

- Connects to Astro2020 Decadal Science Panel Questions E-Q2, E-Q3, E-Q4, G-Q3
- *Related gaps: SCI-15 for effects of stellar photosphere heterogeneities on spectra*

**Gap Summary:**

Systematic effects in time series photometry and high contrast images limit the ability to extract reliable exoplanet spectra amidst backgrounds from residual instrumental signals (detector transients, stellar speckles) or from exozodi. The measured values of empirical parameters such as spectral slopes and linewidths can be affected, and the achieved spectral sensitivity may be worse than the photon noise limit. Early detections of molecular spectral features via transit spectroscopy did not withstand reanalysis (e.g., Deming & Seager 2017, JGRP, 122, 53).

Analysis to establish detection limits can be considered precursor science, while work to refine software or data-taking strategies would be preparatory science. Analysis of flight data to improve algorithms would be follow-up work.

**Capability Needed:**

- Ability to reliably extract physical parameters, such as the atmospheric pressure-temperature profile and abundances of major atmospheric constituents.
- Quantify the effects of uncertainties in exoplanet parameters on the derived atmospheric parameters, toward the goal of defining parameter measurement accuracy requirements. This is especially important to drive mass measurement precision requirements for gap SCI-08.
- Ability to model and subtract spatially varying exozodi backgrounds from exoplanet imaging spectra, and to distinguish the spectral effects of clouds and planetary surface features from the effects of gas phase opacities.
- Thorough understanding of the limits of the data, including effects of correlated and systematic noise sources; and understanding of the limits of spectral retrieval codes through model intercomparisons.
- Strategies for data taking, calibration, and processing to mitigate these issues for each individual instrument/observatory and compilation of lessons learned for future work.

**Capability Today:**

- Community analyses of *JWST*, *HST* & *Spitzer* transit spectra, and of imaging spectra from ground adaptive optics (e.g., GPI, SPHERE).
- Development of best practices for acquiring exoplanet spectra with JWST. Systematics between exoplanet spectra from different JWST instruments have been studied by Carter et al. (2024, Nature Astronomy, 8, 1008).
- Some understanding of the effects of model assumptions on retrieved parameters (Barstow et al., 2020, MNRAS, 493, 4884).
- The ExoPAG SAG 19 report defined new approaches to quantifying detection significance in high contrast imaging datasets.
- Best practices for JWST high contrast imaging were compiled by Hinkley et al. (2023, arXiv:2301.07199).

- The JWST Early Release Science Team for transit spectroscopy held a post-launch data challenge in March 2022 on simulated transit data sets, further information can be found at https://ers-transit.github.io/workshops.html .
- The subtraction of structured exozodiacal cloud emission, needed in order to isolate planetary signals, has been studied by Currie et al. (2024, AJ, 166 197).
- Studies of contamination by stellar photospheric heterogeneities as a limitation to extraction of transiting exoplanet spectra (see gap SCI-15) and stellar speckles as a limitation to extraction of space-based imaging spectra of exoplanets (e.g., Rizzo et al., 2018, SPIE, 10698).
- A data challenge for ground-based high contrast imaging is continuing (https://exoplanet-imaging-challenge.github.io/), with recent results including a study of coronagraphic data processing effects on retrieved spectra (Nasedkin et al., 2023, A&A, 678, A41).
- Gap topic was highlighted as a Precursor Science Gap for the NASA ROSES *Astrophysics Decadal Survey Precursor Science* proposal calls.
- A *HWO* spectral retrieval tutorial workshop took place in November 2024.

**Mitigations in Progress:**

- ExoPAG SAG 26 is leading a comparison of different spectral retrieval tools, with the goal of understanding and reducing systematic discrepancies between them. They will conduct a blind retrieval challenge focused on reflected light spectra of terrestrial exoplanets.

## 2.4. SCI-04: Planetary system architectures: occurrence rates for exoplanets of all sizes

- Connects to Astro2020 Decadal Science Panel Questions E-Q1, E-Q2, E-Q3
- Connects to PSA2022 Decadal Strategic Research Questions 12.2, 12.3, 12.4
- Related gaps: *SCI-05 Occurrence rates and uncertainties for temperate rocky planets* ($\eta_\oplus$)

**Gap Summary:**
The architectures of planetary systems are important for defining the range of outcomes of the processes of planet formation and evolution, and for setting the context for conditions of exoplanets in the habitable zone. Measurements of the distribution of planetary parameters (e.g., number of planets, their masses, and radii; their orbital elements and orbital orientations w.r.t. the stellar rotation axis and stellar companion orbits), from various techniques for stars of various types (e.g., stellar mass, multiplicity, metallicity, evolutionary state) are important both for constraining planetary system formation and evolution models, and for predicting science yields for future NASA missions. The lack of integrated exoplanet population studies limits our understanding of exoplanet demographics over a wide range of planet masses, radii, and orbital separations. Extrapolations to HZ demographics need to be done from the most complete information available (see gap SCI-05).

This is a precursor science gap in terms of informing science yield models for formulating new missions (e.g., *HWO*), a preparatory science gap to enhance the science return on current and future NASA missions (e.g., *Roman*), and a follow-up science gap on capturing exoplanet data from a mission and combining the results from other relevant missions and techniques (e.g., *Kepler*, *TESS*).

**Capability Needed:**
- Planet formation and population synthesis models are needed for comparison against the observed demographics, to improve our understanding of planet formation and evolution, and to provide predictions for still-undetected planet populations that may be detectable with future facilities and observations. The latter need to be in a form that can be readily used to synthesize "universes of exoplanets" for mission concept yield simulations (e.g., *HWO*).
- Integrated exoplanet demographic results from different methods (e.g., transit, direct imaging, RV, and microlensing surveys).
- Include effects of completeness & reliability of *Kepler* detections, the low yield of direct imaging detections of self-luminous planets, microlensing results from recent campaigns.
- Update periodically to include new surveys (e.g., *TESS*) and methods (e.g., astrometry with *Gaia*) and to correct the host star properties used in prior studies.
- Extend the temporal baselines of RV and transit surveys to discover longer-period planets. Constraints on the properties of astrometric perturbers identified from *Gaia-Hipparcos* accelerations.
- Predictions of dynamical interaction effects that may be seen in the distributions of semi-major axes, eccentricities, masses, etc. across an ensemble of planetary systems

- The effect of measurement uncertainties on the occurrence rate results must be quantified. Practitioners of each technique should make sufficient occurrence rate metadata available for later combined analyses (Christiansen et al., 2023, ExoPAG SIG 2 report, arXiv:2304.12442).
- Quantify the impact of stellar binarity on exoplanet frequency, as many potential direct imaging target stars are in multiple systems.

**Capability Today:**

- Ongoing microlensing, RV, transit, and direct imaging projects continue to build statistics on exoplanet frequency distribution (see review by Gaudi, Christiansen, Meyer, 2021, ExoFrontiers, p.2-1, and references therein).
- Community efforts to follow up accelerations of nearby stars calculated by combining *Gaia* & *Hipparcos* datasets, with e.g., direct imaging (e.g., Currie et al., 2023, Science, 380 198).
- A small number of studies have attempted to quantify the impact of stellar binarity on exoplanet frequency (e.g., Moe & Krautter 2021, MNRAS, 507, 5393).
- Research on improving knowledge of the demographics of planets orbiting nearby stars was identified as a Precursor Science Gap for the NASA ROSES *Astrophysics Decadal Survey Precursor Science* proposal calls, and proposed research on the topic was supported.

**Mitigations in Progress:**

- There is a large ongoing community effort to validate *TESS* exoplanet candidates.
- Ongoing community efforts for assessing occurrence rates for close-in planets using *Kepler*, *K2*, and *TESS* data, reconciling results from different discovery methods (e.g., transit, radial velocity, microlensing, direct imaging) and factoring in *Gaia* stellar data.
- *HWO*'s Solar System in Context working group, and its sub-group Demographics & Architectures was tasked with synthesizing current knowledge on exoplanet occurrence rates and architectures for *HWO* target stars, and assessing *HWO*'s sensitivity and working angle range for characterizing exoplanetary architectures of *HWO* target stars.
- ExoPAG SIG 2 is monitoring available data on exoplanet occurrence rates.
- Astrometry: *Gaia* project will be releasing planet candidates in *Gaia* DR4 (December 2026) and DR5 (not before end of 2030). Recent analysis by Lammers & Winn (2025, A.J. 171 18) predict ~7,500 exoplanet discoveries in Gaia DR4 (5-year dataset) and ~120,000 in Gaia DR5 (10-year dataset).
- ALMA studies of the structure of protoplanetary disks, and high contrast imaging searches using both ground-based telescopes and JWST for self-luminous exoplanets.
- *Roman Space Telescope* microlensing survey will measure occurrence rates for the cold planet population.

*Cross-Divisional Synergy*:
Planetary science research on modeling the formation of solar system planets and other small bodies, and on the timing of their formation and migration.

## 2.5. SCI-05: Occurrence rates and uncertainties for temperate rocky planets ($\eta_\oplus$)

- Connects to Astro2020 Decadal Science Panel Questions E-Q1, E-Q2, E-Q3
- Connects to PSA2022 Decadal Strategic Research Questions 12.2, 12.3, 12.4
- Related gaps: SCI-04 (Planetary system architectures: occurrence rates for exoplanets of all sizes)

**Gap Summary:**

A critical parameter guiding the design of the *HWO* recommended by Astro2020, which must be capable of spectrally characterizing habitable zone (HZ) planets orbiting nearby stars, is *eta-earth* ($\eta_\oplus$) - the occurrence rate of rocky exoplanets in the habitable zones of FGK stars. Gap SCI-05 is the subset of gap SCI-04 focusing specifically on frequency of Earth-sized ($R_P \approx$ 0.8-1.4 $R_{Earth}$) planets in HZs. $\eta_\oplus$ remains considerably uncertain, with determinations from different authors varying by almost an order of magnitude. Better characterization of $\eta_\oplus$ will reduce uncertainty in estimated science yields (detection, spectroscopy) and reduce the risk that *HWO* might not achieve the Astro2020 Decadal goal of spectrally characterizing ~25 potentially habitable exoplanets. Measurements of trends in $\eta\oplus$ as functions of stellar parameters (e.g., mass, multiplicity, metallicity, etc.) could improve predictions of yield estimates.

This is a precursor science gap affecting the final architecture of *HWO* and its starlight suppression system. It is follow-up science of *Keple*r and *TESS*, and additional constraints may come from *Roman* microlensing survey.

**Capability Needed:**

- Observations, archival data analysis, and supporting theoretical research enabling improved constraints on $\eta\oplus$, reducing uncertainty and potential biases.
- Detections of temperate rocky planets, and observations which can confirm the existence of candidate temperate rocky planets in *Kepler* data upon which $\eta_\oplus$ critically relies.
- Analysis of occurrence rates taking into account final *Kepler* products and improved stellar parameters, such that remaining uncertainties are dominated by intrinsic *Kepler* systematics. Ideally the values would be constrained and cross-checked via datasets other than *Kepler,* and trends sought as a function of system properties (e.g., stellar mass, multiplicity, presence of larger planets, etc.) to improve the fidelity of yield estimates.
- Development of an error budget for $\eta\oplus$, identifying which terms contribute the most uncertainty and would be ripe for further improvement.

**Capability Today:**

- Published analyses by several authors, including (e.g., Burke et al., 2015, ApJ, 809, 8; Traub 2016, arXiv:1605.02255; Hsu et al., 2019, AJ, 158, 109; Pascucci et al., 2019, ApJ, 883, L15; Bryson et al., 2020, AJ, 159, 279; Kunimoto & Matthews 2020, AJ, 159, 248). Gaia results improved the estimates of radii for all transiting planets (e.g., Fulton & Petigura 2018, AJ, 156, 264; Berger et al., 2018, ApJ, 866, 99).

- The ExoPAG SAG 13 final report informed the *LUVOIR* and *HabEx* mission concept studies, which adopted $\eta_\oplus = 0.24^{+0.46}_{-0.16}$ for yield calculations, a factor of 3 systematic uncertainty.
- The most recent estimates from Bryson et al. (2021, AJ, 161, 36), which informed the Astro2020 Decadal Survey, have 68% confidence limits spanning $0.16 < \eta_\oplus < 1.5$ depending on assumptions about habitable zone and extrapolation of completeness for the final *Kepler* data. Bergsten et al. (2022, AJ, 164, 190) used updated stellar parameters informed by Gaia, final *Kepler* data products and candidate reliability, and attempted to account for the effects of atmospheric loss shaping the small planet population, yielded an occurrence rate estimate of $\Gamma_\oplus = 0.15^{+0.06}_{-0.04}$ ($\eta_\oplus \approx 0.09 \pm 0.03$).
- Members of ExoPAG SIG 2 (Exoplanet Demographics) published a review on the topic of eta-Earth $\eta_\oplus$ and the challenges of improving its measurement (Fernandes et al. 2025, PASP, 137, 121001). A *Kepler*-centric review on the measurement of $\eta_\oplus$ was published by Bryson et al. (2025, PASP, 137, 124401).
- Research on improving knowledge of $\eta_\oplus$ was identified as a topic in the Precursor Science Gap for the NASA ROSES *Astrophysics Decadal Survey Precursor Science* proposal calls, and proposed research on the topic was supported.

**Mitigations in Progress:**

- The community is actively working on planet occurrence rate studies that incorporate final *Kepler* data and *Gaia*.
- While *TESS* was not designed for exoplanet demographic surveys, the community is actively working on determining occurrence rates of terrestrial planets from *TESS*, particularly around M dwarfs.
- The *Roman* Galactic Bulge Time-Domain Survey may provide some constraints on $\eta_\oplus$, however the anticipated results may not come soon enough to inform *HWO* design.

*Cross-Divisional Synergy:*

Improve understanding of the evolution of Venus and Mars to help inform limits on where habitable planets may be found orbiting other stars (i.e., empirical constraints on habitable zone).

## 2.6. SCI-06: Yield estimation for exoplanet direct imaging missions

- Connects to Astro2020 Decadal Science Panel Question E-Q4
- *Related gaps: SCI-05 (Occurrence rates and uncertainties for temperate rocky planets, $\eta_\oplus$)*

**Gap Summary:**
The survey for temperate rocky exoplanets in ~100 cumulative nearby habitable zones, called for by Astro2020, will be the largest single observing program of the *HWO* mission. An accurate definition of this survey will allow the Decadal goal of characterizing ~25 temperate rocky exoplanets to be achieved while preserving mission time for other priority science programs. Quantitative exoplanet science metrics, focusing on the yields of spectroscopically-characterized temperate rocky planets but also including aspects of comparative planetology, are needed to facilitate architecture trades for *HWO*.

This is a precursor science activity for *HWO* mission before Mission Concept Review, with subsequent refinements being preparatory science up through mission launch.

**Capability Needed:**
- Capability to calculate multiple exoplanet science metrics, especially yield estimates, for exoplanet direct imaging missions using an open-source code for the simulator and with provision for community code contributions. The simulator should be implemented independent of mission architecture advocates, in support of the technology maturation program for *HWO*.
- Calculate mission yields and their uncertainties so that the ability of each architecture option to achieve the Decadal goal of characterizing ~25 potentially habitable worlds is understood, as well as the yields of other planet types.
- Community consensus is needed on the key astrophysical and instrument performance parameter inputs to the simulator (gaps SCI-04, SCI-05, SCI-11) and definitions of the exoplanet science metrics to be used. Ideally the effects of sources of contamination, either known measured sources or unknown sources that can be statistically modeled, are included and the impacts of their effects on survey science yields assessed (these may include, e.g., light contributed by exozodiacal light, additional planets, neighboring stars, etc.).
- Improved treatment of observation scheduling and mission rule optimization, in the use of precursor observations in planning imaging observations, and in estimating the significance of planet signals.

**Capability Today:**
- Adaptive Yield Optimization code employed by *HabEx* and *LUVOIR* concept studies (Stark et al., 2019, JATIS 5 4009), and *HWO* (Stark et al., 2024, JATIS, 10, 034006).
- Public ExoSIMs code developed under the *WFIRST* Preparatory Science program by Savranksy & Garrett (2016, JATIS, 2, 011006), and applied in independent analysis of *LUVOIR* and *Habex* yields by Morgan et et al. (2019, Proc. SPIE, 1111701) and *HWO* yield simulations which varied instrument parameters (Morgan et al., 2024, JATIS, 130925M). Morgan et et al. (2022, Proc. SPIE, 121802) presented ExoSIMS yields for 6m space telescope for coronagraph-only, starshade-only and hybrid architectures for several metrics, and accounting for possibility of knowledge from EPRV preparatory science surveys.

- Imaging detection metrics (e.g., Jensen-Clem et al., 2018, AJ, 155, 19 and SAG 19 final report).
- *Bioverse* (Bixel & Apai, 2021, AJ, 161, 228) is a publicly available code for generating planets, simulating surveys and hypothesis testing.
- *ExoVista* (Stark, 2022, AJ, 163, 105) produces synthetic planetary systems and disks and simulates their physical parameters as a function of time.
- Simplified performance assumptions (coronagraph detection metrics, scheduling of starshade observations vs. planet orbital phase and L2 formation-flying dynamics) limit the accuracy of the results.
- Research on yield estimation for *HWO* was listed as a Precursor Science Gap for the NASA ROSES *Astrophysics Decadal Survey Precursor Science* proposal calls, and proposed research on this topic was supported.
- ExEP organized the "Exoplanet Yield Modeling Tools Workshop" at AAS242 in Albuquerque in June 2023 to inform the community of, and encourage community development of, yield modeling capabilities. Recorded talks and tutorial materials were posted online[8].
- *HWO*'s Exoplanet Science Yield Working Group (ESYWG) was chartered to "*provide multiple independent estimates of scientific capability of HWO's exoplanet imaging instrumentation…*" to inform trade studies for instrumentation (including coronagraph). ESYWG published their findings (Stark et al. 2025, arXiv:2502.18556).

**Mitigations in Progress:**

- HWO START teams submitted draft science cases for the observatory in 2025, and the recently formed CSIT (Community Science and Instrument Team) is reviewing them to decide on the science cases that will drive HWO's architecture.

[8] https://science.nasa.gov/astrophysics/programs/exep/resources/exoplanet-meetings/exoplanet-yield-modeling-tools-workshop/

## 2.7. SCI-07: Intrinsic properties of known exoplanet host stars

- Connects to Astro2020 Decadal Science Panel Questions E-Q1, E-Q2, E-Q3, E-Q4
- Connects to PSA2022 Decadal Science Question 12.6d, 12.10
- *Related gaps: SCI-12 (Improvement of knowledge of transiting planet radii)*

**Gap Summary:**

The accuracies of measured exoplanet parameters needed for planetary characterization and interpretation of atmospheric spectra rely directly on the fidelity of stellar parameters derived from photometry, spectroscopy, astrometry, etc. Examples of these parameters are the stellar luminosity, effective temperature, activity indicators, chemical abundances, etc. If an exoplanet is a likely candidate for spectroscopy, then a deep set of stellar parameters will be needed to support interpretation of the spectra. If the exoplanet is unsuitable for spectra (host star or planet is too faint; planet does not transit but still is too close to resolve via imaging) then only a smaller set of stellar parameters would be needed.

This is a follow-up activity once a confirmed planet is found by any detection technique. This is currently follow-up science on discoveries from Kepler/K2 and TESS, which can take advantage of opportunities afforded by observations at a variety wavelengths by other observatories (e.g., *HST, Chandra*, etc.), and constitutes preparatory science for *JWST*, *Ariel*, and *Roman*.

*Note:* Improving knowledge of exoplanet radii, especially for deblending the contributions from stellar companions (both physical and unphysical), is covered separately in gap SCI-12.

**Capability Needed:**

- Improved observational constraints on exoplanet and host star properties are needed to help inform the modeling of exoplanet atmospheres and interpretation of exoplanet spectroscopy (gap SCI-02). Stellar luminosity, age, high energy emission (e.g., UV, X-ray, flare properties), and stellar mass loss/stellar wind properties, help inform modeling of the evolution of the planet (and its star) and habitability studies. Time series observations of variable magnetic activity indicators (e.g., chromospheric activity, UV, X-ray) may be needed to constrain average values, which may also help with constraining age.
- Knowledge of chromospheric activity and EUV emission inform models of atmospheric escape, and measurement of NUV and FUV emission informs modeling of atmospheric photochemistry.
- Precision stellar elemental abundances provide proxy information for the composition of a system's protoplanetary disk and the composition of exoplanet interiors, i.e., a starting point for modeling[9].
- For terrestrial exoplanet studies in the near term, accurate elemental abundances and ages for M dwarf host stars are needed but have proved challenging.

[9] Constraints on the composition of well characterized transiting small planets show that some measures like water mass fraction and Fe/Mg ratio can vary widely for rocky exoplanets compared to host stars, while iron-to-silicate mass ratios for rocky planets vs. their stars show statistical correlation (Adibekyan et al. 2024, A&A, 692, A67).

- Basic stellar parameters (e.g., evolutionary status, mass, metallicity, etc.) are needed for non-exoplanet host stars to enable statistical studies.
- Measurements of the evolution of stellar parameters tied to magnetic activity (e.g., high energy flux, wind, etc.) for age-dated samples (whether exoplanet hosts or not) can provide empirical calibrations for the past history of these stellar parameters to assist modeling of planetary atmospheres.
- Improved knowledge of planetary system architecture, including stellar, substellar, or planetary companions, is helpful for interpretation of exoplanet properties and modeling.

**Capability Today:**

- NASA Exoplanet Archive[10] (NEA) contains compilation of confirmed and candidate exoplanets and their host stars, which can inform mission concept studies focusing on studying transits or transit spectroscopy/ photometry of previously known exoplanets, or direct imaging of previously known exoplanets. NEA entries for exoplanet host stars are cross-IDed to Gaia DR2 and DR3 catalogs, each of which contain astrometry, photometry, spectroscopy data for stars.
- The Hypatia Catalog Database[11] compiles stellar chemical abundance data for thousands of stars including mission target stars and >1300 exoplanet host stars.
- Improved rotation-age-activity trends for cool main sequence stars can be used to age-date exoplanet host stars (e.g., Bouma et al., 2023, ApJ, 947, L3; Engle 2024, ApJ, 960, 62).
- Observations in UV with *HST*, and X-rays with *Chandra* and *XMM-Newton* have provided measurements of high-energy spectra across a broad range of stellar types and ages/activity levels for exoplanet host stars to inform the modeling of exoplanet atmospheres (e.g., France et al., 2016, ApJ, 820, 89).
- Stellar wind properties are challenging to measure, but some observational constraints come from detections of astrospheres in Lyman α absorption with *HST* (e.g., Wood et al., 2021, ApJ, 915, 37) and X-rays from charge-transfer between wind ions and ISM neutrals with *XMM* (e.g., Kislyakova et al., 2024, Nature Astro., 8, 596), among other methods (e.g., Lyman α absorption for exoplanet transits, etc.).
- *CUTE* (*Colorado Ultraviolet Transit Experiment*)[12] cubesat launched in 2021 and its 4-yr mission is measuring NUV transit spectra of close-in transiting planets to constrain exoplanet mass-loss rates and atmospheric composition.
- The ExoPAG SAG 17 final report reviewed the observational needs re: stellar characterization for *TESS* candidates (Ciardi et al., 2018, arXiv:1810.08689).
- The ExoPAG SAG 22 final report listed sets of stellar properties and data that should be obtained, cataloged, maintained, improved, and curated for exoplanet host stars, including the targets of future missions (Hinkel et al., 2021, arXiv:2112.04517).
- The 2025 NExScI conference "Know Thy Star, Know Thy Planet 2" presented the recent state-of-the-art in terms of how stars affect our ability to discover and characterize exoplanets[13].

---

[10] NASA Exoplanet Archive: https://exoplanetarchive.ipac.caltech.edu/ (Christiansen et al., 2025, PSJ, 6, 186).
[11] Hypatia Catalog: www.hypatiacatalog.com (Hinkel et al., 2014, AJ, 148, 54)
[12] *CUTE*: https://lasp.colorado.edu/cute/ (France et al., 2023, AJ, 165, 63)
[13] Know Thy Star, Know Thy Planet 2 Conference: https://conference.ipac.caltech.edu/knowthystar2/

**Mitigations in Progress:**

- NASA Exoplanet Archive is actively compiling data on exoplanets and their host stars.
- *SPARCS* (*Star-Planet Activity Research Cubesat*)[14] is launched in early 2026 to monitor NUV and FUV emission and variability for low-mass stars of a wide range of ages.
- The *UVEX* (*Ultraviolet Explorer*)[15] mission intends to conduct an all-sky survey that should reach 50-100 times deeper than that of the *GALEX* mission of 20 years ago. *UVEX* is expected to launch in 2030 and should provide measurements of the near- and far-UV continuum emission for many more exoplanet host stars.

[14] *SPARCS*: https://sparcs.asu.edu/ (Shkolnik et al., 2025, JATIS, 11, 042212)
[15] *UVEX*: https://www.ipac.caltech.edu/project/uvex

## 2.8. SCI-08: Mitigating stellar jitter as a limitation to sensitivity of dynamical methods to detect small temperate exoplanets and measure their masses and orbits

- Connects to Astro2020 Decadal Science Panel Questions E-Q1, E-Q2, E-Q3, E-Q4, G-Q3
- Connects to PSA2022 Decadal Science Question 12.11
- Related gaps: *SCI-09 Dynamical confirmation of exoplanets, measuring their masses & orbits*

**<u>Gap Summary:</u>**
Measurements of masses and orbits are crucial for characterizing exoplanets, and for modeling their spectra and composition. The radial velocity (RV) and astrometry techniques are anticipated to be the primary means of measuring masses in support of *HWO* reconnaissance of exoplanet atmospheres. However, stellar variability from multiple sources over a range of timescales produces challenges for both techniques. Modern precise RV spectrographs can detect super-Earth (5-10 $M_{Earth}$) planets in the habitable zones of Sun-like stars but stellar variability generally dominates the RV uncertainty budget below 1 m/s and precludes the reliable detection of temperate Earth-mass exoplanets (RV amplitude $K \sim 9$ cm/s for a solar analog). Meanwhile, modern astrometry capabilities are only sensitive to giant planets on many-year orbits but the stellar variability signature in astrometric data is expected to be smaller relative to the ~microarcsecond-level amplitudes predicted for exoEarths around the nearest Sun-like stars.

Dedicated surveys on modern Extreme Precision RV spectrographs could provide a mix of preparatory and/or follow-up capability for detecting small and potentially habitable worlds, constraining their orbits and measuring their masses. Steady improvements in how stellar variability is modeled and mitigated will lead to orbits and masses for ever smaller and more distant orbiting exoplanets. The EPRV Working Group report[16] has presented a community pathway on this topic. Advancing these dynamical methods (both EPRV[17] and astrometry) - or improving our knowledge of their practical limitations - is a precursor science topic that can inform the strategy for obtaining the mass/orbit measurements needed for the small exoplanets *HWO* is being designed to study. Improving the ability to measure masses via EPRV can enhance preparatory or follow-up science for *JWST* and *Ariel* as the measuring the masses of even close-in small transiting planets at the sub-meter/second level remains challenging on current instruments.

[16] EPRV WG Final Report (Crass et al. 2021, arXiv:2107.1429)
[17] The recent review by Burt, Dumusque, & Halverson (2025, AR&AA, arXiv:2511.01954, Fig. 3) classified *Precise Radial Velocities (PRV)* with detectable amplitudes in the 1-3 m/s range, *(Extreme) Precise Radial Velocity ((E)PRV)* in the 0.3-1 m/s range (with latest generation instruments), and *Extremely Precise Radial Velocity* in the <0.3 m/s range (capability to be developed to detect temperate super-Earths and Earth-mass planets orbiting Sun-like stars).

Note: Technology needs for EPRV and astrometry are tracked separately as ExEP Technology Gaps[18].

**<u>Capability Needed:</u>**
Clarification on the exoplanet mass strategy for *HWO* by late 2020s and advancement of EPRV and astrometric capabilities. *<u>What</u> mix of mass-measuring capabilities (EPRV and astrometry) - with <u>what</u> instruments - and <u>when</u> - will be able to provide either prior or posterior knowledge of exoplanet masses and orbits to inform HWO targeting for its survey for potentially habitable worlds and other exoplanets, and modeling of the observed reflected-light exoplanet spectra?* This question can be broken down into parts:

- To what extent - and to what accuracy - do we need to measure the masses and orbits of exoplanets that will be imaged and spectrally characterized with *HWO*?
- Can EPRV be advanced sufficiently to enable reliable measurement of Earth-like exoplanets orbiting Sun-like stars (spanning approximately mid-F through K-type main sequence stars) to inform the *HWO* HZ survey?
- Could the *HWO* itself provide sub-microarcsecond astrometry measurements to sufficient accuracy to provide masses of potentially habitable worlds or are other options needed?

For reference, among the nearest, brightest plausible ~100 *HWO* targets for a survey of potentially habitable worlds, the median target is a 5th magnitude Sun-like (~1.0 $M_{Sun}$) star at $d \sim$ 10 pc, and an Earth twin around a solar twin would produce a radial velocity amplitude of ~9 cm/s (independent of distance & brightness, but dependent on the inclination) or astrometric amplitude of 0.3 microarcsecond (independent of inclination).

*<u>Radial Velocity</u>*: RV variability is intrinsic to Sun-like stars at the few m/s level, and higher for active stars (see EPRV WG Final Report). RV detection of Earth analogs requires an uncertainty in the Doppler amplitude of less than 20% of the planetary amplitude which ranges from 5 - 50 cm/s, depending on host star type and planet's location in the Habitable Zone. This drives a requirement on the uncertainties of relative RV measurements to be below ~10 cm/s over time scales of years to decades so that systematic errors do not dominate. Precise requirements on EPRV surveys are also required as the uncertainty on the Doppler amplitude depends on the structure of correlations in measurement errors as a function of time. Cross-instrument comparisons of the latest generation of EPRV spectrographs (e.g., NEID, KPF, ESPRESSO, EXPRES, etc.) should be carried out to determine which design / reduction decisions produce improved RV performance. Major commitments of observing time on telescopes with such spectrographs are needed to test the field's ability to detect Earth-analog planets as this requires long baseline stellar data sets, preferably with sufficient coordination to ensure that a subset of stars are targeted by multiple EPRV instruments. New analysis methods are needed to correct for stellar variability in RV data using high spectral resolution and broad spectral coverage.

ExoPAG SAG-8 (Plavchan et al., 2015; arXiv:1503.01770) assessed capabilities and future potential for precision radial velocity for exoplanet detection and characterization (e.g., measuring bulk densities of transiting exoplanets) relevant to NASA missions. Extreme Precision Radial Velocity Working Group Final Report (Crass et al., 2021) provided a modern roadmap to

[18] https://science.nasa.gov/astrophysics/programs/exep/technology/

NASA and NSF for advancing the EPRV technique in support of a direct imaging space telescope (i.e., *HWO*).

*Astrometry*: Predicted astrometric amplitudes for 1 $M_E$ planets at the Earth Equivalent Instellation Distance (EEID) for plausible *HWO* target stars are mostly between 0.1-1 microarcsec (Mamajek & Stapelfeldt 2024, arXiv:2402.12414). For Sun-like activity levels, astrometric jitter would be ~0.05 microarcsec – small, but not negligible (and higher for more active stars; Meunier & Lagrange 2022, A&A, 659, A104).

## Capability Today:

*Radial Velocity*: The state of the art in RV detection of exoplanets is summarized in the recent review by Burt, Dumusque & Halverson (2025, ARA&A, in press, arXiv:2511.01954). Fischer et al. (2016, PASP, 128, 066001) presented a previous summary on single measurement precisions (SMPs) and ongoing RV surveys, and updates from the community were presented at the EPRV 5 (2023, Santa Barbara) and EPRV6 (2025, Porto) conferences. Thus far, the smallest reported RV amplitude attributed to an exoplanet, without period and phase info from associated transits, is for Barnard's star e (22.1$\pm$3.8 cm/s; Basant et al. 2025, ApJL, 982, 1), a signal corresponding to a 0.193$\pm$0.033 $M_{Earth}$ planet on a 6.7 day orbit detected using a joint analysis of MAROON-X and ESPRESSO observations. Among red dwarf stars, multiple other planets have been recently reported with RV amplitudes of $K \leq$ 50 cm/s orbiting the two nearest examples: Proxima Centauri (planet d; $K$ = 39 cm/s; Suárez Mascareño et al. 2025, A&A, 700, A11), and three others orbiting Barnard's star (planets b, c, d; $K$ = 44, 45, 43 cm/s, respectively; Basant et al. 2025). Around Sun-like stars, a $K$ = 50 cm/s amplitude exoplanet with orbital period 648 days has been reported around the nearby G star HD 20794 (82 Eri; Nari et al., 2025, A&A, 693, A297). Stellar data from ESPRESSO demonstrate an on-sky RV precision of 40 cm/s over 3.5 years (Figueira et al. 2025, A&A, 700, A174) and stellar data from HARPS-N demonstrate a Doppler sensitivity of 54 cm/s on periods up to 1 year (Anna John et al., 2023, MNRAS, 525, 2) and HARPS-N solar data demonstrated the feasibility of reliably recovering RV signals down to ~20 cm/s amplitude for the Sun for orbital periods shorter than ~900 days (Dumusque et al. 2026, A&A, 706, A231).

Numerous advancements in RV data acquisition and analysis techniques show promise in reducing the impact of stellar variability on simulated and real solar and stellar datasets:

- Availability of data and software: high cadence, high SNR solar observations are being produced by multiple instruments and going public early. More RV data reduction pipelines and post processing pipelines are being made open source. An EPRV standardized data format is in development.
- Dedicated queue scheduling, already in place at WIYN and Gemini, and Keck is transitioning that direction.
- Simulations of stellar spectra and impacts of some types of stellar variability with SOAP GPU (Zhou & Dumusque 2023, A&A, 671, A11) and GRASS (Palumbo et al., 2024, AJ, 168, 46)

- Use of machine learning techniques to model / mitigate stellar variability (e.g., de Beurs et al., 2022, AJ, 164, 2; Liang et al., 2024, AJ, 167, 1; Ford et al., 2024, arxiv:2408.13318).

Note that stars hotter than mid-F and/or with high *v*sin*i*, which makes up tens of percent of the nearby direct imaging targets (e.g., Mamajek & Stapelfeldt 2024, arXiv:2402.12414), may not have the Doppler information required for EPRV measurements and astrometry may be required to detect / characterize temperate rocky planets around these stars.

*Astrometry*: Studies of stellar astrometric jitter for the Sun-as-a-star and stars of varying activity levels for the following missions and concepts: *SIM-Lite* (Makarov et al., 2010, ApJ, 717, 1202), *Theia* (e.g., Meunier & Lagrange, 2022, A&A, 659, A104), *Gaia* and *Small-JASMINE* (e.g., Sowmya et al., 2021, ApJ, 919, 94). Existing ground-based astrometry (CHARA, NPOI, VLTI) cannot reach the required accuracy.

**Mitigations in Progress:**

- *HWO* Technology Maturation Project Office has Working Groups providing input on a measurement needs for preparatory and follow-up exoplanet orbit and mass information from EPRV and/or astrometry that will be imaged and spectrally characterized with *HWO*, and conducting analyses on the feasibility of sub-microarcsecond astrometry using *HWO* High Resolution Imager.
- *NASA ROSES:* NASA funded a second EPRV Foundation Science grant opportunity in ROSES 2022, and the program was continued and expanded as the new ROSES element D.20 "*Exoplanet Mass Measurement Program*" (EMMP) in ROSES 2024.
- *Precision Radial Velocity Spectrographs*: Major NASA investment in NEID EPRV instrument (NEID) on the 3.5-m WIYN (northern hemisphere 3.5-m observatory) through NN-EXPLORE. NEID has been operating since 2021 (stability demonstrated to <30 cm/s) as is available to the community via NOIRLab with proposals reviewed by a special Exoplanet TAC. Data for NEID RV standard stars[19] as well as continuous observations of the unresolved disk integrated sun are made public immediately, via NExScI[20]. Other recent, non-NASA, EPRV instruments include EXPRES on the 4.3-m Lowell Discovery Telescope, KPF on the 10-m Keck I telescope, ESPRESSO on the 8-m Very Large Telescope, and MAROON-X on the 8-m Gemini North telescope. Most instruments are demonstrating RV precision at the 30-50 cm $s^{-1}$ level over timescales of months. US community members can access KPF via NASA Keck time proposals, MAROON-X via NOIRLab proposals, and ESPRESSO via ESO proposals. EXPRES is not currently available for general user proposals but the EXPRES disk integrated solar RV data will be made available via NExScI starting in 2026.
- *EPRV RCN*: Following recommendations from the Exoplanet Science Strategy (2018) and EPRV WG report, and in support of the Astro2020 Decadal scientific vision to "*identify and characterize Earth-like extrasolar planets*" ("*Pathways to Habitable Worlds*"), and to scientifically support future use of *HWO* to fulfill its Decadal goal to search for habitable zone planets and search for biosignatures, ExEP is sponsoring the

[19] NEID Standard Star list: https://www.wiyn.org/Instruments/wiynneid_observers.html
[20] NEID Archive: https://neid.ipac.caltech.edu/search.php

*EPRV Research Coordination Network*[21]. The EPRV RCN aims to increase communication and collaboration in the radial velocity community in order to advance the EPRV technique towards the goal of detecting temperate, small planets around Sun-like stars. The RCN includes over 100 members, and has regular virtual meetings and a colloquium series on topics like instrumentation, observations/surveys, data analysis techniques (e.g., stellar variability mitigation techniques, etc.), solar studies, etc.

- Queloz et al. are deploying the new HARPS3 spectrograph to the 2.5 m Issac Newton telescope in 2026 for a high-cadence, long-term monitoring campaign on the nearest sun-like stars, an effort known as the Terra Hunting experiment[22].

[21] https://science.nasa.gov/astrophysics/programs/exep/nnexplore/eprv/eprv-rcn/
[22] https://www.terrahunting.org/

## 2.9. SCI-09: Dynamical confirmation of exoplanet candidates and determination of their masses and orbits

- Connects to Astro2020 Decadal Science Panel Questions E-Q1, E-Q2, E-Q3, E-Q4
- Connects to PSA2022 Decadal Science Question 12.11
- Related Gaps: *SCI-08 (Mitigating Stellar Jitter)*

**Gap Summary:**

The majority of current exoplanet discoveries have been made via the transit method, e.g., *Kepler*, *K2*, *TESS*. However, transit observations typically do *not* constrain the planetary mass (except for rare cases where transit-timing variations [TTVs] can), which is crucial for understanding the planetary bulk density / composition and interpreting atmospheric spectra. RV observations usually need to be made to obtain mass measurements, but do not provide orbital inclinations. Orbital ephemerides need to be known precisely enough to support scheduling of transit and eclipse spectroscopy.

This is a follow-up activity for exoplanets not discovered via the radial velocity method.

**Capability Needed:**

There are insufficient precision RV resources available to the community to follow up all *K2* and *TESS* candidates that may be relevant to spectroscopic studies with *JWST* and *Ariel*. Follow up of *Kepler/K2* and *TESS* candidates starts with quick-look low-precision RV screening for false positives (e.g., eclipsing binaries), then high precision PRV to determine masses and orbital elements of the best candidates. *TESS* follow-up requires RV observing time in N and S hemispheres, sufficient to cover the expected 12.5k *TESS* candidates of which ~1,250 should be detected in the 2-min cadence data, with ~18 smaller than 2 $R_{Earth}$ and in the habitable zone (Kunimoto et al., 2022, AJ, 163, 290). Continuation of the *TESS* extended mission, and/or targeted transit followup with other facilities, to refine transit ephemerides sufficiently for predicted transit time accuracy better than 1 hour through the epochs of the *Ariel* mission. Dynamical follow-up of *Gaia* astrometrically-detected planet candidates after the DR4 and DR5 data releases, in order to refine their orbits and to inform future imaging surveys. Measurements of orbital inclinations with respect to stellar rotation axes via the Rossiter-McLaughlin effect to constrain the planetary system's dynamical history.

**Capability Today:**

*TESS* has achieved the main goal for its primary mission of detecting ~50 exoplanets smaller than Neptune and measuring their masses. *TESS* continues operations in its extended mission.

- *RV:* Follow-up of *TESS* targets is ongoing with RV facilities including NEID, KPF, HPF, MAROON-X, Magellan/PFS, HARPS, HARPS-N and ESPRESSO for precise follow-up at ~1 m/s precision. ESPRESSO has demonstrated RV precision of ~28 cm/s over a night and ~50 cm/s over several months for HD 85512, with instrument precision of ~10 cm/s, and there is ESPRESSO-GTO survey of ~50-100 *K2/TESS* planets (<2 $R_{Earth}$, V < 14.5 mag; Pepe et al., 2021, A&A, 645, 96). For instruments with demonstrated <1 m/s RV accuracy, there is limited US community access, and only in the northern hemisphere (e.g., NEID, KPF, MAROON-X, etc.).

- *TTV*: e.g., analysis of *Kepler* multi-planet systems; *Spitzer* Exploration Program Red Worlds campaign observed transits over 1000+ hrs for 7-planet TRAPPIST-1 system (Ducrot et al., 2020, A&A, 640, A112, and Agol et al., 2021, PSJ, 2, 1), *CHEOPS* is now contributing TTV observations (e.g., Vivien et al., 2023, A&A, 688, A192).

**Mitigations in Progress:**

- NASA-NSF Partnership for Exoplanet Observational Research (NN-EXPLORE). NN-EXPLORE supports community access to the NEID instrument on Kitt Peak, as well as instrument maintenance.
- Options for renewed southern hemisphere community PRV access are being explored. In the past, NN-EXPLORE supported US community access to SMARTS 1.5-m CHIRON and MINERVA-Australis.
- NASA supports community access to Keck HIRES and KPF instruments (which includes queue-based scheduling).
- The *TESS* mission has been extended for the third time starting in 2025 following the Astrophysics Senior Review of Operating Missions. *TESS* Follow-up Observing Program (TFOP)[23] is continuing to observe and analyze data for new *TESS* exoplanet candidates.

[23] https://tess.mit.edu/followup/

## 2.10. SCI-10: Observations and analyses of direct imaging targets

- Connects to Astro2020 Decadal Science Panel Questions E-Q1, E-Q2, E-Q3, E-Q4
- Related Gaps: *SCI-07 (Properties of Exoplanet Host Stars)*

**Gap Summary:**
For the targets of high contrast imaging missions, precursor and preparatory observations fall into two categories. First, for direct imaging mission *science target stars*: 1) screening for close-in, low-mass stellar and substellar companions that might compromise exoplanet imaging sensitivity; 2) detecting exoplanets for future characterization, or setting observational and/or dynamical limits on their presence; 3) measuring stellar physical properties, chemical abundances, and radiation environments to enable accurate planet characterization including interpretation of exoplanet spectra (see gap SCI-07); and 4) identifying systems with high exozodi levels, in the form of both warm and hot dust, where spectroscopy of small exoplanets may not be possible (see gap SCI-11). The second target category is that of *PSF reference stars* for use as coronagraph point spread function (PSF) calibrators for direct imaging missions. Identifying a sample of bright, single PSF calibrator stars with small angular diameters is especially important for setting up coronagraphic “dark holes” via iterative wavefront control.

Precursor science on likely *HWO* science and PSF reference star targets includes screening them for faint companions; precision radial velocity and astrometry time series observations; and improving the observational limits on exozodi levels. For the *HWO* science targets, precursor science also includes obtaining improved orbital elements for the multiple systems; and dynamical simulations of habitable zone stability for stars with known stellar, substellar, and planetary mass companions. For these targets, measurements of their stellar physical & chemical properties, and their radiation environments, is *HWO* preparatory science that may be completed in the years leading up to launch.

**Capability Needed:**
Working versions of target catalogs for the *HWO* and *Roman* Coronagraph missions, for both science targets and PSF reference stars. A census and characterization of plausible *HWO* target systems, including their stellar and substellar companions which may dynamically limit the presence of planets in their habitable zones. Theoretical research constraining the stability of planets in the habitable zones of *HWO* targets. Assess the bound companion (stellar and substellar) detection limits provided by existing data (e.g., RV, *Gaia* astrometry, high contrast imaging, etc.). With the goal of detecting temperate rocky exoplanets in the target systems and other planets that may affect the dynamical stability of planet orbits in habitable zones, conduct PRV observing programs in both N and S hemispheres (executed consistently over > 5 years, ideally through *HWO* launch and survey), and conduct observations using other techniques which may feasibly detect small planets orbiting nearby target stars (including e.g., astrometry, and high-contrast IR imaging). Constraints on stellar multiplicity from high resolution imaging, RV and astrometry (e.g., *Gaia*), are needed for both science target stars and PSF reference stars to assess whether high contrast imaging will be feasible, as starlight suppression performance is affected by the presence of close neighboring stars. Uniform determination of stellar properties across the target sample in both hemispheres. Sufficiently sensitive X-ray and far-UV characterization of stellar radiation environments in the target systems. Screening for the presence of bright exozodi clouds. Identification of bright PSF calibrator stars spread around the sky, screened to show no close companions or disk nebulosity, stars which the *Roman* Coronagraph and *HWO* could use to set up coronagraphic "dark holes" just prior to observing a science target. The priority of preparatory research on potential *Roman* Coronagraph science targets is unclear, as a decision is still pending on whether to conduct post-tech-demo science surveys with the instrument. The ExoPAG SAG 22 report includes recommended datasets to complete host star characterization.

**Capability Today:**

- *Catalogs*: Catalogs of target stars that are bright enough and near enough to enable a survey for exo-Earths with a range of plausible *HWO* architectures have been released in recent years. The ExEP preliminary HWO target catalog (Mamajek & Stapelfeldt 2024, arXiv:2402.12414) now comprises the "Tier 1" list of 164 HWO target stars, and two lower tier classes of target stars have been defined (Tuchow et al., 2025, PASP, 137, 104402). The HWO Preliminary Input Catalog (HPIC) of ~13k stars from Tuchow, Stark & Mamajek (2024, AJ, 167, 139), is large enough to support simulations of science yields for larger exoplanet types in more distant systems.
- *Knowledge of companions*: The census of even massive companions (low-mass stars, brown dwarfs, giant planets) is still incomplete for plausible *HWO* targets, let alone for small exoplanets. The *HabEx* and *LUVOIR* teams conducted yield simulations based on the Hipparcos star catalog with limited binary star info and initial estimates of key stellar astrophysical parameters. Binary orbits in the nominal target systems are often poorly characterized, including several cases where *Gaia* was unable to confirm companions reported in the Washington Double Star Catalog[24]. Astrometric accelerations for likely *HWO* targets have been summarized by Painter et al. (2025, AJ, 170, 147). Near-IR interferometry with VLT/Gravity or CHARA could place limits on close companions in systems not amenable to precision radial velocity techniques. A preliminary assessment of the multiplicity of bright (Vmag<3), small angular diameter (<2 mas) candidate PSF reference stars for Roman Coronagraph Instrument and HWO has been conducted (Hom et al., 2026, A.J. 171 36), downselecting to a list of 58 stars requiring further vetting.
- *Radial Velocity Surveys*: The NASA/NSF EPRV Working Group recommended a strategy for a precursor observing program. ExEP supported two archival PRV studies of nearby Sun-like stars, many of which are likely to eventually be *HWO* target stars (Howard & Fulton 2016, PASP, 128, 4401; Laliotis et al., 2023, AJ, 165, 176). Observational constraints on the presence of companions in a given mass and orbital period space of HWO target stars have been published (e.g., Rosenthal et al., 2021, ApJS, 255, 8; Harada et al., 2025, AJ, 170, 343). Kane et et al. (2024, AJ, 168, 195) conducted a dynamical study of 30 potential *HWO* target stars already known to harbor exoplanets and found that for 11 of them, known giant planets already exclude >50% of their habitable zones as dynamically viable for temperate small planets.
- *High Energy Emission:* Peacock et al. (2025, AJ, 170, 293) present a comprehensive survey of the available archival X-ray and UV observations for HWO Tier 1 target stars.

[24] https://www.astro.gsu.edu/wds/

Some analysis of far-UV and X-ray observations for plausible *HWO* target stars has been conducted thus far (e.g., Binder et al., 2024, ApJS, 275, 1).

- *Exozodiacal dust levels*: The LBTI HOSTS survey measured exozodi levels in ~20 likely *HWO* targets (Ertel et al., 2020, AJ, 159, 177). Knowledge of exozodi levels is discussed further in gap #11.
- *Facilities*:
  - Radial velocity instruments: e.g., NEID, Keck HIRES and KPF, Lick APF, HARPS, HARPS-N, PFS-Magellan, EXPRES, MAROON-X, HARPS3, etc.
  - High contrast imaging: Keck, Gemini, ExEP-supported speckle program, etc. JWST campaigns on the very nearest stars are capable of reaching thermal emission at modest zodi levels (~5-10x) and giant exoplanets (e.g., α Cen A; Beichman et al. 2025, ApJL, 989, L22). VLT/NEAR observations of α Cen A by Wagner et et al. (2021, Nature Comm. 12, 922) demonstrates limits of IR sensitivity to giant exoplanets around nearest target stars.
- *NASA support*: Gap topic was highlighted as a Precursor Science Gap for two NASA ROSES *Astrophysics Decadal Survey Precursor Science* proposals calls (2022, 2023).

**Mitigations in Progress:**

- The *HWO* Target Stars & Systems sub-group of Living Worlds Working Group will report on the state of the art for stellar system data for potential *HWO* target systems, and anticipated science data needs for *HWO* exoEarth and exoplanet survey work for plausible *HWO* target stars. This WG has task groups investigating data availability and future needs for various types of stellar data needed to interpret exoplanet spectra (e.g., stellar high energy emissions, multiplicity, fundamental parameters, etc.).
- Precision radial velocity surveys: NEID GTO program on WIYN is surveying ~20% of ExEP preliminary HWO target list stars, and the EXPRES GTO program on LDT is surveying ~10-15% of NASA Mission Targets.
- *Gaia* DR3 detected accelerations versus Hipparcos results for a large number of stars (e.g., Kervella et al., 2022, A&A, 657, 7; Feng et al., 2022, ApJ, 262, 21), suggesting the presence of low-mass companions, some of which may be good targets for direct imaging detection and characterization.
- *Gaia* DR4 (anticipated release: December 2026) is expected to reveal astrometric perturbations by giant exoplanets for thousands of stars (Lammers & Winn, 2025, A.J. 171 18), the closest and brightest of which may be good targets for direct imaging detection from the ground and space.
- Many of the *HWO* target stars are being searched for close stellar companions by optical speckle imaging.
- Pre-launch screening of candidate PSF reference stars for contaminating companions to enable coronagraph dark hole calibration for Roman via a series of ground-based high resolution imaging projects by the *Roman* CPP Team are ongoing. Use of the *Roman* Coronagraph to screen candidate HWO reference stars for faint companions will be considered by the *Roman* CPP Team as a potential survey project after the mission tech demo objectives have been met.

## 2.11. SCI-11: Understanding the abundance and distribution of exozodiacal dust

- Connects to Astro2020 Decadal Science Panel Questions E-Q1, E-Q2, E-Q3, E-Q4
- Related gaps: *SCI-03 (Spectral signature retrieval)*, *SCI-06 (Mission yield simulations)*

**Gap Summary:**
Exozodiacal dust is a noise source that compromises imaging and direct spectroscopy of small planets in and around the habitable zones of nearby stars. It can be much brighter in a telescope resolution element than exoplanetary signals, especially at longer wavelengths. Substructure in the exozodi distribution may mimic the presence of an exoplanet and thus confuse searches made with smaller telescope apertures, or of more distant targets. To date, substructure in the distribution of habitable zone dust has been mapped only for the case of our own solar system.

Reducing the uncertainty in the median and distribution of exozodi levels and addressing the risk that the presence of hot dust may affect scattered light levels in the HZ, are precursor science activities. Other activities cited here are preparatory science.

**Capability Needed:**
Statistical knowledge of exozodiacal dust levels in the habitable zone relative to the level in our solar system is needed for nearby FGK stars that will be the targets of HWO. Further observational work that reduced the uncertainty in current determinations of the median exozodi level would reduce the risk that HWO's survey of nearby habitable zones would consume too large a fraction of the prime mission, that the mission lifetime might need to be increased, or that the mission might produce lower-quality spectra than intended. Theoretical modeling of dust sources & sinks, dust transport processes, and dynamical sculpting by planets. Mission yield simulations of how exozodi levels and uncertainties affect the integration times and achievable signal-to-noise ratios for exoplanet detection and characterization, as a function of mission architecture. Simulations of scenes as viewed by future imaging missions, quantifying the effectiveness of multi-epoch observations to discriminate exozodi clumps from planets. Directly observed scattered light images of exozodi disks in the habitable zones of HWO targets would be very valuable, if they were sensitive down to the ~10 zodi level, were obtained for stars with measured 10 μm excess (potentially enabling dust albedo estimates), and had the resolution to show substructures and validate theoretical simulations. An understanding of the physical relationship (if any) between the hot dust emission detected by near-IR interferometers and the population of small grains in habitable zones.

**Capability Today:**
Images are available showing the substructure of cold (Kuiper Belt) debris disks as seen by *HST*, ground adaptive optics, Herschel, and ALMA. There is a rich literature of theoretical models of debris disk structure treating such effects as dust radial transport and planetary perturbations on debris disk structure. Hot dust emission is detected in many systems with near-IR interferometry but its origin, relevance to habitable zone dust, and potential risk to detection of temperate rocky exoplanets is not understood. At least one reported strong detection of hot variable dust turned out to be most likely due to an unseen low-mass stellar companion (Stuber et al. 2026, AJ, 171, 1). The LBTI HOSTS survey searched for mid-IR excess emission due to warm exozodiacal dust in the habitable zones of 38 stars (Ertel et al., 2020, AJ, 159, 177), deriving a median exozodi level 3 times that of the solar system for the subsample of FGK stars, but with a significant +1σ uncertainty of 6 zodis. Detection upper limits for individual FGK stars are only ~120 zodis (3σ), however. While the yields of exoplanet direct imaging missions are a weak function of the exozodi level (Stark et al., 2015, ApJ, 808 139; Stark et al., 2024, JATIS, 10 4006), the quality of spectra of Earth analogs can become problematic for the more distant targets if the exozodi level is > +1σ from the LBTI median result.

**Mitigations in Progress:**

- The Astro2020 Decadal Survey was silent on whether additional investments in exozodi measurements should be a priority. Options for further observational work include:
    - The LBTI instrument team studied possible upgrades that would increase the sensitivity of their instrument by up to a factor of ~3.
    - *Roman* Coronagraph Science Investigation Team published a paper (Douglas et al., 2022, PASP, 134, 024402) quantifying the sensitivity the instrument might be able to achieve to exozodiacal dust in a survey of nearby stars, should sufficient performance and observing time be available with that instrument.
    - Current near-IR interferometers and upcoming ELTs will have capabilities to constrain warm exozodi levels and these are still being assessed.
- ExoPAG SAG 23 is working on a multi-chapter community report on the Impact of Exozodiacal Dust on Exoplanet Direct Imaging Surveys, with completion expected in early 2026. The report is expected to include findings on the best ways to make theoretical and observational progress in exozodi and debris disk studies.
- The first VLTI/MATISSE measurements are providing L band visibility constraints on hot dust. A new nulling interferometer under development for the VLTI "NOTT" will provide higher S/N detections of hot exozodiacal dust in the L band circa 2025.
- NASA XRP is funding theoretical studies and observational efforts to connect various observables.
- Research on "Understanding the Abundance and Distribution of Exozodiacal Dust" was identified as a gap topic in the Precursor Science Gap for the NASA ROSES-2022 call *Astrophysics Decadal Survey Precursor Science*, and proposed research on the topic was supported.

*Cross-Divisional Synergy:*

Research on interplanetary dust grains, understanding source dust populations and distribution of dust in the solar system.

## 2.12. SCI-12: Measurements of accurate transiting planet radii

- Connects to Astro2020 Decadal Science Panel Questions E-Q1, E-Q2, E-Q3, E-Q4
- Related gaps: *SCI-07 Properties of Exoplanet Host Stars*

**Gap Summary:**

Accurate measurements of exoplanet radii are important for: properly characterizing exoplanets, estimating bulk densities, modeling their compositions, atmospheres, spectra, and the discovery of trends important to understanding planet formation and evolution. The accuracy of transiting exoplanet radii is most often limited by the accuracy of measured stellar radii, which can be dominated by the blending effects from neighboring stars. For small exoplanets (i.e., with shallow transits), planet radii accuracy is limited by low S/N, and degeneracies between star-planet radius ratio, limb darkening, and transit impact parameter. Not accounting for light contamination by companions or neighboring stars, or poor stellar characterization, can lead to exoplanet radii systematically miscalculated at the tens % level. AO and speckle imaging validation of *Kepler* prime mission candidates took ~3 years and was only ~70% complete. *TESS* has exceeded the number of *Kepler/K2* candidates and continues to find planets in the extended missions and thus needs a similar long-term effort to determine precise exoplanet radii. The orbital periods for single transit planet candidates can be estimated using the density of the star and the transit parameters, hence improvements in knowledge of stellar densities can improve estimates of orbital periods and the efficiency of transit and radial velocity follow-up campaigns.

This is follow-up science for *Kepler/K2*, and *TESS*, while also being preparatory science to vet targets for observations of exoplanets with *JWST* and *Ariel*.

*Note:* Improving the knowledge of other stellar parameters to help inform the interpretation and modeling of exoplanet data (e.g., spectra) is covered separately in gap SCI-07.

**Capability Needed:**

Detailed observations are needed to derive accurate stellar radii, including high resolution spectroscopy, high resolution imaging, and seeing-limited photometry to identify false positives (e.g., variable stars). High resolution imaging to identify stellar companions is necessary to validate thousands of *TESS* candidates and to prioritize systems for detailed mass determination and atmospheric characterization. For those stars, access to observatories equipped with AO or speckle imaging cameras and reduction pipelines, is needed in both N and S hemispheres. Support work that improves estimation of stellar and exoplanet parameters for discovered exoplanet systems. Supporting photometric and spectroscopic stellar data, along with astrometric, photometric, and spectroscopic data from latest *Gaia* data releases, are critical for accurately assessing stellar parameters – and exoplanet radii. High resolution spectroscopy can reveal spectroscopic binaries, and can provide precise stellar parameters, particularly when coupled with *Gaia* data and high resolution imaging. Injection and recovery tests can place further quantitative constraints on companions. In some cases, asteroseismic analysis of light curves can improve estimates of the star's density, improving estimates of $a/R$, improving constraints on transit radius ratio (along with limb darkening and impact parameter). For occurrence rate studies, accurate limiting radii for planet detection for transit survey stars for which transiting planets were not detected is also important. See ExoPAG SAG 17 report discussing resource needs for *TESS* follow-up to constrain stellar and exoplanetary radii.

**Capability Today:**

NESSI speckle camera at WIYN, and Zorro and 'Alopeke cameras on Gemini S and N, respectively, and NIRC2 on Keck, offer ability to screen a subset of targets to very small separations. Other community resources include SOAR HRCam (speckle), and various ground-based AO observations with e.g., Hale/P3K+PHARO, Robo-AO, VLT/NACO, etc., have helped validate KOIs, K2 candidates, and TOIs. *Gaia* photometry & astrometry resolves well-separated multiples and provides parallaxes that have greatly reduced the uncertainty in intrinsic stellar radii and high resolution imaging resolves close-in multiples beyond the reach of *Gaia*. For improving knowledge of host star $T_{eff}$, metallicity, gravity: high-resolution spectroscopy surveys (e.g., California-*Kepler* survey), lower resolution spectroscopy surveys (e.g., APOGEE & LAMOST), and community access to spectrographs for extracting stellar spectra (e.g., Keck HIRES, NEID, CHIRON, etc.). In some cases for nearby bright exoplanet host stars, interferometry can be used to measure stellar diameters.

**Mitigations in Progress:**

- Ongoing NASA support for community access to optical speckle cameras on WIYN, Gemini-N and Gemini-S, as well as near-IR AO imaging with Keck/NIRC2.
- Community seeing-limited and high contrast imaging observations, and high resolution spectroscopy, supporting *TESS* follow-up.
- SPHEREx measurements of late-type exoplanet hosts will provide more accurate photometry and thus improved stellar radius estimates.
- ExoFOP is supporting community work in this area through coordination of observations, and the sharing of data and derived results.

## 2.13. SCI-13: Properties of atoms, molecules and aerosols in exoplanet atmospheres

- Connects to Astro2020 Decadal Science Panel Questions E-Q1, E-Q2, E-Q3, E-Q4
- Connects to PSA2022 Decadal Science Question 12.6
- Related Gaps: SCI-16 Biosignatures
- *See gap SCI-02 for atmosphere modeling issues beyond the properties of its constituents*

**Gap Summary:**
Understanding and interpreting the full gamut of exoplanet spectra emerging from *JWST* & ground telescopes, and to come from the *Roman* Coronagraph Instrument, *Ariel*, and *HWO* hinges on the ability to link observations to theoretical atmosphere models. These models rely on understanding the optical properties of atoms, molecules and aerosols, as well as the reaction rates between relevant chemical species. Together these properties shape our understanding of the chemistry and climate of exoplanet atmospheres. For some species in some wavelength or temperature regimes (e.g., $CO_2$ in the near-UV), line lists or emission/absorption strengths are imprecisely known and this limits the ability to model exoplanet spectra and interpret biosignatures.

This is precursor science supporting atmosphere modeling which can define *HWO* spectroscopy requirements, especially in the near-UV, and can be follow-up science if observed spectra show new or poorly-modeled features.

**Capability Needed:**
Ability to perform theoretical calculations of key molecular and atomic spectroscopic properties in relevant physical conditions including effects of pressure broadening. A non-exhaustive list of challenges includes:

- Obtaining lab measurements or performing ab initio calculations of line intensities, line positions, pressure or collisional broadening, and partition functions.
- Ability to perform theoretical calculations and/or laboratory measurements of gas spectra, reaction rate coefficients, and aerosol properties in relevant physical conditions.
- Ability to obtain refractive indices of aerosol properties in relevant physical conditions.
- Additional measurements of Archean-analog haze optical constants are needed.
- The UV cross-sections for even common gases may be poorly constrained in modern literature, but may be of critical importance for modeling early planet photochemistry (when stars have more UV emission, and oxygen and ozone may not be present to efficiently absorb stellar UV). Measurements, improvements, and validation of UV (especially >200 nm) absorption cross-sections of photochemically important molecules and biosignature gases (e.g., $CO_2$, $H_2O$, $CH_4$[g]; all of which have been identified as underconstrained with observable implications in the literature (e.g., Ranjan et al., 2020, ApJ, 896, 148; Broussard et al., 2024, ApJ, 967, 114). Needs for improving the fidelity of synthetic exoplanet spectra for *HWO* formulation include e.g., independent validation of $H_2O$ cross-sections for temperate conditions, and improvement measurements of $CO_2$ cross-sections for temperate conditions.

- Greater emphasis on laboratory work for molecules and aerosols relevant to the early Earth but not present in the modern Earth atmosphere.

See white papers by Fortney et et al. (2016; arXiv:1602.06305) and Savin et et al. (2019, BAAS 51, 3, 96) for summaries on this topic.

**Capability Today:**

Ab initio line list calculations of several dozen molecules with the ability to correct line positions. Laboratory measurements of line lists at low temperatures and in the near-UV. Reaction rate coefficients measured at high combustion temperatures and standard Earth temperatures. Curated exoplanet aerosol database of refractive indices (provided by HITRAN) over limited wavelength ranges. Several publicly available databases of line lists and opacities relevant for exoplanets exist, for example:

- HITRAN[25] (*hi*gh-resolution *tran*smission) molecular absorption database
- 2024 release of ExoMol[26] database of molecular line lists for planetary atmospheres
- Molecules and Atoms in (Exo)planet Science: Tools and Resources for Opacities (MAESTRO)[27] database
- HELIOS-K 2.0 Opacity Calculator and Open-source Opacity Database for Exoplanetary Atmospheres hosted at DACE[28]

**Mitigations in Progress:**

- Several exoplanet specific efforts continue to expand the parameter space and accuracy of line list data (e.g., HITRAN/HITEMP, ExoMol, MAESTRO, TheoReTS, XRP proposals).
- The HITRAN 2024 update will be released in December 2025, accompanied by a JQSRT special issue.
- Funded collaboration between HITRAN/ExoMol and exoplanet theory groups to develop community tools and best practices for computing and disseminating opacity data.
- XRP-supported programs on measuring spectroscopic line lists, absorption cross-sections, etc., for common molecules in atmospheres of hot planets and brown dwarfs relevant to impending *JWST* observations, and for modeling warm terrestrial planets relevant to both *JWST* observations and predicting spectra for *HWO*.
- Gap topic was highlighted as a Precursor Science Gap for the NASA ROSES-2022 call *Astrophysics Decadal Survey Precursor Science*.

[25] https://hitran.org/
[26] https://www.exomol.com/ (Tennyson et al., 2024, JQSRT, 326, 109083)
[27] https://science.data.nasa.gov/data-sites/opacities
[28] https://dace.unige.ch/opacityDatabase/ (Grimm et al., 2021, ApJS, 253, 30)

## 2.14. SCI-14: Exoplanet interior structure and material properties

- Connects to Astro2020 Decadal Science Panel Questions E-Q1, E-Q2, E-Q3, E-Q4
- Connects to PSA2022 Decadal Science Question 12.5, 12.7
- Related gaps: *SCI-02 (Modeling exoplanet atmospheres)*

**Gap Summary:**

Improved understanding of interior structure across the mass/radius diagram would be valuable for interpreting the results of current and future exoplanet missions. Exoplanets exhibit a wide range of densities beyond those seen in the solar system. Tenuous "super puffs", extremely dense planets, and the wide range of radii observed over the mass range of ~2-10 $M_E$ and their trends with orbital radius, all pose challenges to models of exoplanet structure, composition, formation, and evolution. The application of planetary interior, formation, and evolution models is hampered by uncertainties in the measured or numerically-predicted material properties under the relevant physical conditions, including the solubility of gases which can be exchanged with the atmosphere, and including conditions experienced transiently during planet formation.

This is largely follow-up science to model the interiors of planets with mass, radius, and/or atmospheric composition measurements.

**Capability Needed:**

Experimental measurements and theoretical calculations of material properties (e.g., equations of state, transport properties, mixing properties, etc.) under the pressure-temperature conditions found in exoplanets (super-Earths, sub-Neptunes, and cores of giant exoplanets). Development of dynamic compression experiments that simulate the pressure and temperature conditions in deep interiors. Access to material properties data (e.g., phase relationships, thermodynamic properties) in an organized format and robust modeling tools over a wide range of pressures, temperatures, and compositions (e.g., $H_2O$, $NH_3$, $CH_4$ ice, silicate-ice mixtures, silicate-hydrogen mixtures, H/He, etc.). Progress in understanding how variations in a rocky planet's bulk silicate composition can affect mantle convection, surface geochemistry, and the availability of trace elements needed for surface biology.

**Capability Today:**

Facilities to explore Earth-interior pressures already exist. Ab initio computer simulations exist but not with broad application. Only limited data exist beyond solar system planet conditions. Theoretical modeling of exoplanet interior evolution, suite of models across varying planet density, internal heat flux, starting compositions (e.g., Thorngren et al., 2016, ApJ, 831, 64; Lopez & Fortney 2014, ApJ, 792, 1; Dorn et al., 2018, ApJ, 865, 20). NASA-supported workshops like the NExSS/NAI/NSF Joint Workshop "Upstairs Downstairs: Consequences of Internal Planet Evolution for the Habitability and Detectability of Life on Extrasolar Planets" (2016), "Habitable Worlds 2017: A System Science Workshop", and NExScI-supported "Exoplanet Demographics" (2020).

**Mitigations in Progress:**

- NASA ROSES XRP supports investigations to explore the chemical and physical processes of exoplanets (including state and evolution of surfaces, interiors, and atmospheres).
- NASA Astrobiology/ICAR includes support for research on how volatiles are exchanged between the atmosphere, surface, and interior of exoplanets.

*Cross-Divisional Synergy*:
Support for NASA Earth Science research on constraining interior composition and structure of Earth, and Planetary Science research on the same for the terrestrial planets (e.g., InSight), gas giants (e.g., Juno), and ice giant planets.

*Cross-Agency Synergies:*
Center for Matter at Atomic Pressures (CMAP)[29] is a new NSF Physics Frontier Center designed to connect observational and laboratory scientists to address the high pressure microphysics relevant to exoplanetary interiors.

[29] https://cmap.rochester.edu/

## 2.15. SCI-15: Quantify and mitigate the impacts of stellar contamination on transmission spectroscopy for measuring the composition of exoplanet atmospheres

- Connects to Astro2020 Decadal Science Panel Questions E-Q1, E-Q2, E-Q3, E-Q4, G-Q3
- Related Gaps: *SCI-03 (Spectral signature retrieval)*

**Gap Summary:**
Transmission spectroscopy is an important method for probing the composition and structure of the atmospheres of exoplanets, and for the next two decades will be the primary means of studying the atmospheres of small rocky exoplanets. The method relies on time-series stellar spectroscopy, monitoring wavelength- dependent brightness variations. However, the stellar disk that the planet crosses is heterogeneous, with spatial and temporal variations in emission from the photosphere, chromosphere, spots, faculae, and plages - whose effects can conspire to mimic some molecular band features expected for exoplanets. To exploit observations with current (e.g., *HST, JWST*) and future (e.g., *Ariel, HWO*) space observatories, we need to be able to quantify the degree of stellar contamination and develop mitigation strategies.

This is follow-up science to understand the limits of exoplanet transmission spectroscopy data for individual stars.

**Capability Needed:**
The ExoPAG SAG 21 report[30] presents detailed needs regarding the stellar contamination issue for advancing the transmission spectroscopy technique. Theoretical and observational research on time-varying properties of spots, faculae, and granules in vis/IR (0.3-5μm) for the Sun and exoplanet host stars of varying type and activity. Include MHD modeling to provide priors on spectra of spots and faculae for stars of varying parameters, validation of 3D granulation simulations against vis/IR observations of Sun and other stars. Long time baseline panchromatic observations of Sun and stars to inform models of the stellar atmospheres and its heterogeneities, and synthetic spectra (unresolved disk, and transit chords). Best practices are needed for including stellar heterogeneity into transit modeling and atmospheric retrievals. A consensus needs to emerge on the practical limits to the detection of exoplanet spectral features, and which exoplanet spectral features are the least susceptible to stellar contamination.

**Capability Today:**
See SAG 21 report (Sec. 2) for extensive review on current capabilities. Observations of spot occultations from ground and space (e.g., *Kepler*) have enabled joint analyses of the properties of the transiting planet and star and its surface phenomena. A JWST transmission spectrum of TRAPPIST 1-b, a planet thought not to have an atmosphere and thus be a useful fiducial for measurement of stellar noise levels (Lim et al., 2023, ApJ, 955, L22), shows stellar contamination signals an order of magnitude larger than the photon noise limit, starkly demonstrating the need to develop mitigation approaches.

[30] Rackham & Espinoza et al. (2022, arXiv:2201.09905; id. 2023, RASTI, 2, 148) doi.org/10.1093/rasti/rzad009

The SAG 21 report lists numerous publicly available starspot occultation codes and recent work on the subject. Some recent examples include: Herbst et et al. (2021, ApJ, 907, 89) presents updated empirical relations for starspot temperature and area for FGKM stars. Johnson et et al. (2021, MNRAS, 504, 4751) fit *Kepler* light curves of variable cool stars using models of time-varying faculae and spots and employing 3D magnetoconvection models. Barclay et et al. (2021, AJ, 162, 300) demonstrated that starspot contamination could plausibly account for the claimed detection of $H_2O$ for K2-18b. Rackham & de Wit (2024, AJ, 168, 82) explore the use and limitations of using out-of-transit stellar spectra and synthetic spectra to model stellar contamination, and propose future work to advance this topic.

**Mitigations in Progress:**

- Ongoing JWST programs taking transmission spectra of exoplanets around low-mass stars, further quantifying the nature and extent of contamination signals, and acting as a proving ground for mitigation methods.
- The *Pandora* smallsat launched in January 2026 and will collect simultaneous visible photometry and NIR spectroscopy data for transiting exoplanets of K/M dwarf stars, allowing the effects of spots along the transit chord to be assessed (Quintana et al., 2024, Proc. SPIE, 13092, 1309214).
- NSO *DKIST* 4m solar observatory offers high spatial and temporal resolution observations[31] of the solar photosphere and chromosphere (Rimmele et al., 2020, Sol. Phys., 295, 172).
- "*Cambridge Workshops of Cool Stars, Stellar Systems and the Sun*" conferences on related topics are held biennially (next meeting "CS23" will be in Tokyo in 2026).

*Cross-Divisional Synergy:*
Observations of *Sun through Living With a Star* (LWS) Program of the NASA Heliophysics Division.

*Note:* The observational and modeling research on the Sun and stars for gap SCI-15 has a connection to the needs for gap SCI-08 (understanding and mitigating stellar jitter for disk-integrated stellar spectra for precise radial velocity and astrometric measurements).  There may be synergies between the mitigation approaches for RV and spectroscopic noise.

[31] https://nso.edu/dkist/data-center/

## 2.16. SCI-16: Building the inventory of remotely observable exoplanet biosignatures and their false positives

- Connects to Astro2020 Decadal Science Panel Questions E-Q2, E-Q3, E-Q4
- Connects to PSA2022 Decadal Science Questions 12.9, 12.11
- *Related Gaps: SCI-02 (Modeling Exoplanet Atmospheres)*

**Gap Summary:**
There has been significant advancement identifying false positives and false negatives and constraining detectability of atmospheric biosignatures in recent years, renewed interest in the detectability of surface biosignatures, and fast growth in explorations of novel technosignatures as signs of technological life. Further work in these frontier topics would advance our ability to search for biosignatures, interpret them in the context of their environment, and strengthen their interpretation with regard to the search for life.

**Capability Needed:**
Both the Astro2020 Decadal Survey and 2018 Astrobiology Strategy NAS report recommend that NASA support research on characterizing novel biosignatures, agnostic biosignatures, planetary processes that can change or mimic biosignatures, and develop statistical frameworks for assessing biosignatures taking into account system observables (to assess confidence that potential biosignatures might be due to life). Astro2020 (p. 4-9) states "*theoretical calculations of planetary atmosphere chemistry and evolution will be needed to interpret biosignature gases detected in exoplanet spectra. This theoretical research lays the groundwork for designing new observational programs and planning for new facilities*." More studies are needed to explore the parameter space, constrain and define self-consistent star-planet-life reference case scenarios for detectability of biosignatures, and solidify frameworks for quantifying uncertainty in inference of habitability and the presence of life.

Understanding observable biosignatures for potentially habitable planets is important 1) preparatory science to inform the interpretation of data from *JWST* observations of transiting small temperate exoplanets, and 2) precursor science for *HWO t*o inform instrument choices (e.g., wavelength coverage, spectral resolution, S/N requirements, etc.) to enable it to carry out the Decadal-envisioned exo-Earth survey campaign ("*to search for biosignatures from a robust number of about ~25 habitable zone planets"*). The wavelength limits of *HWO* high contrast imagine (from near-UV through near-IR) and spectroscopy capability of HWO should factor in biosignatures, including the challenging case of whether high contrast near-UV observations of the ozone feature near ~0.25 µm in exoplanet spectra can be supported. As Earth was habitable for billions of years early in its history, but oxygen-poor (with correspondingly weak $O_2$ and $O_3$ spectral absorption), biosignatures for worlds analogous to Proterozoic and Archean Earth are of particular interest.

Further research on the following topics could enhance our ability to design missions and interpret data from spectra of potentially habitable worlds:

- Potential biosignatures need to be understood in their proper photochemical context, including photochemical predictions of which biosignatures or photochemical false positives might be detected for a given planet orbiting a star of given spectrum (i.e., spectral type and its UV spectrum).
- Spectral line lists for novel biosignatures (e.g., methylated gases, species in high pressure environments).
- Surface reflectance biosignatures (e.g., pigmented microorganisms, vegetation red edge, spectral flux content for light harvesting pigment adaptations, circular polarization), and false positives for these relative to mineral spectra.
- Temporally varying biosignatures.
- Analyses of biosignatures for alternative biochemistries ("Life as We Don't Know It", LAWDKI) and technosignatures could expand the scope of the search for life outside the solar system with current and future facilities beyond focusing on the search for planets like Earth, its past stages, and predicted future stages.

Supporting capabilities needed on cross-sections for important biosignature gases are compiled under Gap SCI-13: Properties of atoms, molecules and aerosols in exoplanet atmospheres.

**Capability Today:**

- The state of the biosignatures field is reviewed by Schwieterman & Leung (2024, Rev. Miner. & Geochem., 90, 465), in Ch. 4 of the NAS Astrobiology Science Strategy report, and in the series of 6 papers in the June 2018 issue of *Astrobiology*[32].
- Krissansen-Totten et et al. (2025, arXiv:2507.14771) present an analysis of atmospheres of rocky planets and potentially habitable worlds to constrain the UV and near-IR limits required by *HWO* for detecting signs of life with reflected light spectroscopy.
- "*Community Report from the Biosignatures Standards of Evidence Workshop*" from NfoLD/NExSS Standards of Evidence for Life Detection Community Workshop (2021)
- *JWST* transit spectroscopy observations of transiting temperate rocky exoplanets orbiting M dwarfs, but stellar noise (gap SCI-15) is a major obstacle to progress.
- Modeling of individual target systems (e.g., TRAPPIST-1 planets, Proxima Centauri b).
- Survey of potential biosignature gases by Seager, Bains, and Petkowski (2016, Astrobiology, 16, 465).
- *HabEx* and *LUVOIR* studies presented detailed science cases for searching for biosignatures and confirming habitability (e.g., *LUVOIR* Signature Science Case #2) and incorporated it into their strategies for the search for life.
- Oxygen in Planetary Biospheres conference (Green Bank Blumberg Astrobiology Workshop 2023) focused on the rise of oxygen on Earth, and role in the search for life.
- Geochemical research on evolution of Earth: chemistry of its atmosphere/ lithosphere / hydrosphere, the early evolution of life (e.g., Kacar 2024, Ann. Rev. Microbio. 78, 23), oxygenic photosynthesis (e.g., Fischer et al., 2016, AREPS, 44, 647), and comparative studies of the evolution of Earth, Venus, and Mars (e.g., Lammer et al., 2018, A&ARv, 26, 2), etc.

[32] June 2018 issue of Astrobiology with 6 review articles on exoplanet biosignatures: https://www.liebertpub.com/toc/ast/18/6

- The Sagan Summer Workshop 2023[33] theme was "*Characterizing Exoplanet Atmospheres: The Next Twenty Years*" and included topics of biosignatures and technosignatures. Presentations are available online.
- Recent research on LAWDKI and implications for biosignatures was recently summarized by Grefenstette et et al. (2024, Astrobiology, 24, S-186).
- *Technosignatures*: IAU Symposium 404 on this topic took place in March 2026 and the talks are available online at YouTube. 2020 Technoclimes Workshop (Haqq-Misra et al., 2022, Acta Astronautica, 198, 194) summarized future theoretical and observational studies on TS and science cases for future missions. Penn State Extraterrestrial Intelligence Center is sponsoring annual symposia (2022, 2023, 2025).

**Mitigations in Progress:**

- NASA ROSES ICAR opportunities are supporting multiple projects aimed at identifying habitable exoplanet targets or modeling exoplanet biosignatures.
- The *HWO* Project has two science sub-working groups organized under the Living Worlds WG: Biosignature Possibilities and Biosignatures Interpretation.
- Research on modeling biosignatures was listed as a Precursor Science Gap for the NASA ROSES *Astrophysics Decadal Survey Precursor Science* proposal calls, and research on this topic was supported.
- *Technosignatures:* ExoPAG SAG 25 Technosignatures is currently active[34]. Following up on the 2018 Technosignatures Workshop, ExEP is supporting a technosignatures gap study to survey the field and catalog approaches in a systematic manner.

*Cross-Divisional Synergy:*

NASA Astrobiology program and its Research Coordination Networks (RCNs): Nexus for Exoplanet System Science (NExSS), Network for Life Detection (NfoLD), Prebiotic Chemistry and Early Earth Environments (PCE3), Network for Ocean Worlds. NASA Astrobiology Science Conferences (AbSciCons) and Exoplanets in our Backyard meeting series. NASA ROSES element C.4 Habitable Worlds supports research on using "*knowledge of the history of the Earth and the life upon it as a guide for determining the processes and conditions that create and maintain habitable environments*." ROSES element C.5 Exobiology supports research on "*the origin and early evolution of life, the potential of life to adapt to different environments, and the implications for life elsewhere*," with one of the emphases on "*Biosignatures and Life Elsewhere*." A new NASA Decadal Astrobiology Research and Exploration Strategy (DARES) is being developed in 2026 and will include findings on exoplanet biosignatures.

[33] https://nexsci.caltech.edu/workshop/2023/
[34] https://science.nasa.gov/astrophysics/programs/exep/exopag/sags/#sag-25

## 2.17 SCI-17: Understanding planet formation and disk properties

- Connects to Astro2020 Decadal Survey Science Panel Questions E-Q1c, E-Q3a
- Connects to PSA 2022 Decadal Science Questions 12.1 & 12.3
- Related Gaps: *SCI-04 Planetary System Architectures*

**Gap Summary:**

Understanding the physical and chemical evolution of protoplanetary disks provides the essential context for interpreting planets' potential for habitability. The ultimate goal is to understand how the raw disk materials are built into planets, and therefore go from statistical distributions of disk properties seen around different types of stars to setting constraints on where and what type of planets can form within the disks, and by extension the composition of their interiors and atmospheres. It is still challenging to accurately constrain total disk gas masses across the young stellar object population. Estimates of disk density and temperature structures have significant uncertainties, which affects the retrieved chemical abundances. A better understanding of the processes of planetesimal formation, planet migration, and interior differentiation would also allow for the inference of planetary properties that cannot otherwise be directly observed. A better understanding of late-stage accretion and interior outgassing of volatiles would inform the initial conditions of planetary atmospheres which set the stage for the evolution of life.

Improving our understanding of planet formation and disk properties is important for several NASA missions in different ways:

- *HST, JWST, Ariel, HWO*: Enhances the science return of spectroscopic observations of exoplanet atmospheres using both time-series observations and direct observations with these space observatories, by helping to place these planets in context.
- *Roman*: Enhance the science return of the *Roman Space Telescope* by providing theoretical and observational support to interpret the demographics of the new planet population probed through the bulge microlensing survey.

**Capability Needed:**

An improved understanding of the diversity, structures, and chemistry within protoplanetary disks. This will include theoretical and observational constraints on the origins and demographics of disks and their structures. It will also include constraints on how the global properties of disks evolve over time. A firm understanding of whether rings and gaps seen in outer disks with ALMA are connected to embedded planets is critical. Observations which show how the water and organics abundances, and elemental ratios evolve in protoplanetary disks are of particular interest, to try and link to measurements of exoplanet atmospheric abundances. Spectroscopic observations of protoplanetary disks across a diverse range of evolutionary stages, stellar mass, and stellar environments are needed to constrain disk properties and chemical compositions (e.g., Krijt et al., 2022, ASPC, 534, 1031; Oberg et al., 2023, ARA&A, 61, 287). Near- and mid-IR spectra by ground-based telescopes and JWST could provide direct constraints on the chemical compositions at the disk surface layer of the inner few au region. Supplemented with thermo-chemical models, we can trace the mid-plane compositions and the transport of volatiles and organics into the terrestrial planet-forming region over time. Far-IR facilities with sufficient spectral resolution and sensitivity to measure fluxes of the HD (1-0) rotational line emission in a large population of disks would provide much better constraints of gas disk masses. Infrared (5-550µm) spectroscopic observations of water gas lines and ice features are key to tracing cold water components and constraining snowline locations. Detailed characterization of disks that host nascent planets like PDS 70 and WISPIT 2 would provide important test cases on the link between disk properties and nascent planets.

More observational and theoretical constraints on the planet formation process itself. The development of the techniques and capability to identify and characterize nascent planets in disks over a range of stellar host types & evolutionary stages, as well as their circumplanetary material. Constraints on the timescale of planet formation, including its start time and how this can depend upon the initial conditions. A better understanding of theoretical mechanisms of planet formation, including e.g., pebble accretion, streaming instability, migration, etc., and understanding the range of physical conditions for planet formation, including how these conditions depend upon stellar mass and the surrounding star-formation environment. Linking the demographics of disks to those of exoplanets in time is also critical to pin down the main physical processes shaping planetary systems (e.g., Mulders et al., 2015, ApJ, 798). The time evolution of orbital architectures needs further study, so as to link present-day exoplanet demographics and distributions of orbits to the initial conditions in the protoplanetary disk (and vice versa).

**Capability Today:**

- ALMA, *JWST, HST*, and large ground-based surveys to measure protoplanetary disk chemistry and spatial structures. Imaging studies are largely sensitive to the outer parts of protoplanetary systems and to large planets. Scattered light studies are biased towards edge-on systems where starlight suppression artifacts are largely absent. The detailed chemical studies by ALMA are mainly only available for two dozen of the brightest sources.
- Just a few examples (PDS 70, WISPIT 2) of planetary systems with planets that appear to be in the process of accretion, and which host circumplanetary material (Benisty et al., 2021, ApJ, 916, L2; Li et al., 2025, ApJL, 990, 70; Close et al., 2025, ApJ, 990, L9).
- Upper limits to the presence of protoplanets from coronagraphic imaging surveys of young stars, but with a significant ambiguity on how extinction within the disk affects the detection sensitivity (Wallack et al., 2024, AJ, 168, 78).
- Population synthesis models with very simplified chemical processes have been developed to link the disk compositions to the atmospheric C/O ratio in gas giants.
- Some detections of young exoplanets using the transit technique (e.g., Barber et al., 2024, Nature, 635, 574).
- The Great Link conference considered connections between disk chemistry and the chemistry of exoplanetary atmospheres (Feinstein et al. 2025 arXiv 2506.00669)

**Mitigations in Progress:**

- There are approved JWST Cycle 1-3 observations to spectrally characterize the inner regions of more than 100 protoplanetary disks, although the parameter coverage is still limited. Most of the existing observations have been focused on disks around ~0.5-1 $M_{\odot}$ stars between ~1-3 Myr-old.
- The SPHEREx[35] Explorer mission is now operational and taking 1-5 µm low-resolution spectra of every source on the sky. For protoplanetary disks, the large mission dataset should constrain the abundances of polycyclic aromatic hydrocarbons across the early stages of disk evolution, and the presence of ices in molecular clouds and in the subset of disks seen close to edge-on.
- Development of far-infrared mission concepts. The PRIMA[36] probe mission proposal would advance understanding of disk chemistry and is now completing its Phase A study. The Astro2020 Decadal survey recommended an eventual far-IR flagship mission.

[35] https://spherex.caltech.edu/ (Bock et al., 2025, arXiv:2511.02985)
[36] https://prima.ipac.caltech.edu/

# 3. ExoPAG Science Analysis Groups (SAGs)

The list and descriptions of the active and completed ExoPAG Science Analysis Groups (SAGs) is maintained at: https://science.nasa.gov/astrophysics/programs/exep/exopag/sags/

## 3.1 Active SAGs

- SAG 23: The Impact of Exo-Zodiacal Dust on Exoplanet Direct Imaging Surveys (initiated 2022)
- SAG 24: Exploring the Complementary Science Value of Starshade Observations (initiated 2023)
- SAG 25: Technosignatures (initiated 2024)
- SAG 26: Exoplanet Reflectance Spectroscopy for the Habitable Worlds Observatory (initiated 2024)

## 3.2 Completed SAGs

Previously completed SAGs are listed in inverse chronological order. SAGs 3, 4, 6, and 7 were initiated but later deactivated; often their topic areas were assigned to other SAGs.

- SAG 22: Investigating an Exoplanet Target Star Archive (2020)
- SAG 21: Effect of Stellar Contamination on Space-based Transmission Spectroscopy (2020)
- SAG 20: Strategies for Mitigating the Impact of the JWST Delay on Exoplanet Science (2020)
- SAG 19: Exoplanet Imaging Signal Detection Theory and Rigorous Contrast Metrics (2017)
- SAG 18: Metrics for Direct-Imaging with Starshades (2017)
- SAG 17: Community Resources Needed for K2 and TESS Planetary Candidate Confirmation (2017)
- SAG 16: Exoplanet Biosignatures (2017)
- SAG 15: Exploring Other Worlds: Observational Constraints and Science Questions for Direct Imaging Exoplanet Missions (2017)
- SAG 13: Exoplanet Occurrence Rates and Distributions (2017)
- SAG 12: Scientific potential and feasibility of high-precision astrometry for exoplanet detection and characterization (2017)
- SAG 11: Preparing for the WFIRST Microlensing Survey (2014)
- SAG 10: Characterizing the Atmospheres of Transiting Planets with JWST and Beyond (2015)
- SAG 9: Exoplanet Probe to Medium Scale Direct-Imaging Mission Requirements and Characteristics (2015)
- SAG 8: Requirements and Limits of Future Precision RV Measurements (2015)
- SAG 5: Exoplanet Flagship Requirements and Characteristics (2013)
- SAG 2: Potential for Exoplanet Science Measurements from Solar System Probes (2010)
- SAG 1: Debris Disks & Exozodiacal Dust (2012)

# 4. Glossary of Common Acronyms for NASA ExEP

| | |
|---|---|
| A&A | Astronomy & Astrophysics |
| AJ | Astronomical Journal |
| ALMA | Atacama Large Millimeter Array (observatory in Chile) |
| AO | Adaptive Optics |
| AO | Announcement of Opportunity |
| APD | Astrophysics Division |
| APF | Automated Planet Finder (robotic 2.4-m optical telescope at Lick Observatory) |
| ApJ | Astrophysical Journal |
| ApJL | Astrophysical Journal Letters |
| ApJS | Astrophysical Journal Supplement Series |
| APOGEE | Apache Point Observatory Galactic Evolution Experiment |
| APRA | Astrophysics Research and Analysis |
| ARC | Ames Research Center |
| ARIEL | *Atmospheric Remote-sensing Infrared Exoplanet Large-survey* (approved ESA M4 mission, with NASA as partner, targeting 2029 launch), now called "*Ariel*" |
| ASTRA | Astrophysics Strategic Technology and Research Accelerator |
| au | Astronomical Unit (symbol is "au" per IAU 2012 Resolution B2) |
| AURA | Association of Universities for Research in Astronomy |
| AXIS | Advanced X-ray Imaging Satellite |
| BD | Brown Dwarf |
| CGI | (*deprecated*) CoronaGraph Instrument on the *Nancy Grace Roman Space Telescope*. The preferred name is now "Coronagraph Instrument."[37] |
| CHARA | Center for High Angular Resolution Astronomy |
| CHEOPS | CHaracterizing ExOPlanet Satellite (ESA mission) |
| CHIRON | CTIO HIgh ResolutiON spectrometer (instrument on CTIO/SMARTS 1.5-m telescope at Cerro Tololo Inter-American Observatory (CTIO), Chile) |
| CMAP | Center for Matter at Atomic Pressures |
| CME | Coronal Mass Ejection |
| CMF | Core Mass Fraction (ratio of mass of Iron-rich core to total planet mass) |
| CMOS | complementary metal oxide semiconductor |
| CNES | Centre national d'études spatiales (National Centre for Space Studies), French national space agency |
| C/O | Carbon/Oxygen ratio |
| COPAG | Cosmic Origins Program Analysis Group (PAG supporting community coordination and analysis for NASA Cosmic Origins Program) |
| COR | Cosmic ORigins Program |
| CoRoT | Convection Rotation and Planetary Transits (CNES/ESA mission) |
| CPD | Circumplanetary Disk |
| CRN | Chemical Reaction Network |
| CTIO | Cerro Tololo Inter-American Observatory (in Chile) |
| CUTE | *Colorado Ultraviolet Transit Experiment* (CubeSat) |
| DKIST | Daniel K. Inouye Solar Telescope (NSF National Solar Observatory facility) |

[37] https://science.nasa.gov/mission/roman-space-telescope/coronagraph/

| | |
|---|---|
| DMS | Dimethyl Sulfide |
| DPAC | *Gaia* Data Processing and Analysis Consortium |
| DR | Data Release |
| DSCOVR | *Deep Space Climate ObserVatoRy* |
| DST/HCIT | Decadal Survey Testbed/High Contrast Imaging Testbed |
| EC | Executive Committee |
| EEID | Earth Equivalent Insolation Distance (EEID; $a_{EEID} = \sqrt{L}$ au where L is stellar luminosity in solar units, au is astronomical unit) |
| ELT | Extremely Large Telescope (formerly European-Extremely Large Telescope) |
| EMMP | Exoplanet Mass Measurement Program (NASA ROSES element soliciting investigations to identify and mitigate systematics that limit radial velocity or astrometry observations from measuring masses of temperate terrestrial exoplanets orbiting Sun-like stars) |
| EOS | Equation of State |
| EPOS | Exoplanet Population Observation Simulator |
| EPRV | Extreme Precision Radial Velocity |
| eROSITA | Extended Roentgen Survey with an Imaging Telescope Array |
| ERS | Early Release Science (*JWST* program) |
| ESA | European Space Agency |
| ESM | Emission Spectroscopy Metric (see Kempton et al., 2018, PASP, 130, 114401) |
| ESO | European Southern Observatory |
| ESPRESSO | Echelle Spectrograph for Rocky Exoplanets and Stable Spectroscopic Observations (instrument for ESO VLT observatory) |
| ESS | Exoplanet Science Strategy (2018) National Academies Report |
| ESYWG | Exoplanet Science Yield Working Group (for Habitable Worlds Observatory) |
| EUV | Extreme Ultraviolet (wavelength range ~9-91.2 nm, or alternative ~10-121 nm) |
| ExEP | Exoplanet Exploration Program |
| Exo-C | *Exo-Coronagraph* (2015 NASA Probe Mission Study) |
| ExoMol | *Mol*ecular line lists for *Exo*planet and other hot atmospheres (database) |
| Exo-S | *Exo-Starshade* (2015 NASA Probe Mission Study) |
| ExoPAG | Exoplanet Program Analysis Group (PAG supporting community coordination and analysis for NASA Exoplanet Exploration Program) |
| ExoSIMS | Exoplanet Open-Source Imaging Mission Simulator |
| EXPRES | Extreme PREcision Spectrometer (instrument on Lowell Discovery Telescope) |
| ExSDET | Exoplanet Standard Definitions and Evaluation Team |
| FFI | Full Frame Images |
| FGK | Stellar spectral types "F", "G", "K" – bracketing stars with "Sun-like" temperatures between about 3900-7200 Kelvin (Sun is G2 with $T_{eff}$ = 5772K) |
| FIR | Far-Infrared (wavelength range ~25-300 μm) or Far-IR |
| FOV | Field of View |
| FUV | Far UltraViolet (wavelength range ~91.2-200 nm, or alternatively 122-200 nm) |
| GALAH | GALactic Archaeology with HERMES |
| G-CLEF | GMT Consortium Large Earth Finder |
| GCM | General Circulation Model |
| GI | Guest Investigator |
| GMT | Giant Magellan Telescope |
| GO | Guest Observer |
| GOE | Great Oxidation Event |
| GOMAP | Great Observatories Mission and Technology MAturation Program (*deprecated* with transition to Habitable Worlds Observatory Technology Maturation Project Office) |
| GPI | Gemini Planet Imager (instrument built for Gemini South 8.1-m telescope) |
| GSFC | Goddard Space Flight Center |
| GTO | Guaranteed Time Observations |
| HabEx | *Habitable Exoplanet Imaging Mission* (pre-Astro2020 concept study) |
| HARPS | High Accuracy Radial velocity Planet Searcher (instrument on ESO 3.6-m telescope at La Silla) |
| HARPS-N | High Accuracy Radial velocity Planet Searcher-North (instrument on Telescopio Nazionale Galileo 3.6-m telescope, La Palma, Canary Islands, Spain) |
| HATNet | Hungarian-made Automated Telescope Network |
| HD | Henry Draper (star catalog) |
| HEM | High Eccentricity Migration |
| HERMES | High Efficiency and Resolution Multi-Element Spectrograph (instrument on Anglo-Australian Telescope) |
| HET | Hobby-Eberly Telescope |
| H-He | Hydrogen-Helium (mix of gases found in stars & some planetary atmospheres) |
| HIRES | High Resolution Echelle Spectrometer (instrument for W. M. Keck Observatory) |
| HITEMP | High-TEMPerature molecular spectroscopic database |
| HITRAN | HIgh-resolution TRANsmission molecular absorption database |
| HOSTS | Hunt for Observable Signatures of Terrestrial Planetary Systems |
| HPIC | *HWO* (*Habitable Worlds Observatory*) Preliminary Input Catalog |
| HRCam | High-Resolution Camera (speckle instrument on SOAR 4.1-m telescope) |
| HRD | Hertzsprung-Russell Diagram |
| HST | *Hubble Space Telescope* |
| HWO | *Habitable Worlds Observatory* |
| HZ | Habitable Zone |
| IAU | International Astronomical Union |
| ICAR | Interdisciplinary Consortia for Astrobiology Research |
| IMF | Initial Mass Function (of stars and brown dwarfs) |
| IPAC | Infrared Processing and Analysis Center |
| IR | Infrared |
| IROUV | *InfraRed Optical UltraViolet* space telescope (*deprecated*, now *Habitable Worlds Observatory*) |
| IRTF | NASA Infrared Telescope Facility |
| JASMINE | *Japan Astrometry Satellite Mission for INfrared Exploration* (mission concept) |
| JATIS | Journal of Astronomical Telescopes, Instruments, and Systems |
| JGRP | Journal of Geophysical Research: Planets |
| JPL | Jet Propulsion Laboratory |
| JQSRT | Journal of Quantitative Spectroscopy and Radiative Transfer |
| JWST | *James Webb Space Telescope* |
| KOI | *Kepler* Object of Interest |
| KPF | Keck Planet Finder (radial velocity instrument for W. M. Keck Observatory) |
| KPIC | Keck Planet Imager and Characterizer |
| KPNO | Kitt Peak National Observatory (site of WIYN observatory) |
| LAMOST | Large Sky Area Multi-Object Fibre Spectroscopic Telescope |
| LAWDKI | Life As We Don't Know It |
| LAWKI | Life As We Know It |
| LBT | Large Binocular Telescope |
| LBTI | Large Binocular Telescope Interferometer |
| LCIT | Large Mission Concept Independent Assessment Team |
| LCOGT | Las Cumbres Observatory Global Telescope Network |
| LDT | Lowell Discovery Telescope (formerly Discovery Channel Telescope or DCT) |
| LUVOIR | *Large UV/Optical/IR Surveyor* (pre-Astro2020 concept study) |
| LWS | Living With a Star (NASA Heliophysics program) |
| MAESTRO | Molecules and Atoms in (Exo)planet Science: Tools and Resources for Opacities (NASA-supported opacities database https://science.data.nasa.gov/opacities/) |
| MAROON-X | Magellan Advanced Radial velocity Observer of Neighboring eXoplanets (instrument on Gemini-North telescope) |
| MCR | Mission Concept Review |
| MIDEX | Medium-Class Explorer |
| MINERVA-Australis | Miniature Exoplanet Radial Velocity Array – Australis (observatory at Mt. Kent Observatory, Queensland, Australia) |
| MIR | Mid-Infrared (astronomical wavelength range ~3-25 μm) |
| MKO | Maunakea Observatories |
| MMF | Mantle Mass Fraction (ratio of mass of silicate-rich mantle to total planet mass) |
| MNRAS | Monthly Notices of the Royal Astronomical Society |
| MO or MoO | Mission of Opportunity |
| MSFC | Marshall Space Flight Center |
| NACO | Nasmyth Adaptive Optics System (instrument for VLT observatory) |
| NAI | NASA Astrobiology Institute |
| NASA | National Aeronautics and Space Administration |
| NEA | NASA Exoplanet Archive |
| NEID | NN-Explore Exoplanet Investigations with Doppler spectroscopy (pronounced '*noo-id*' – derived from the word meaning 'to see' in native language of the Tohono O'odham, on whose land Kitt Peak National Observatory is located) |
| NEOWISE | Near-Earth Object Wide-Field Infrared Survey Explorer |
| NESSI | NASA Exoplanet Star (and) Speckle Imager (instrument on WIYN observatory) |
| NExScI | NASA Exoplanet Science Institute |
| NExSS | Nexus for Exoplanet System Science |
| NfoLD | Network for Life Detection (NASA Research Coordination Network) |
| NIR | Near-Infrared (wavelength range ~0.7-3 μm) |
| NIRC2 | Near InfraRed Camera 2 (instrument for W.M. Keck Observatory) |
| NN-EXPLORE | NASA-NSF EXoPLanet Observational Research |
| NOAO | National Optical Astronomy Observatory (*deprecated,* now part of NOIRLab) |
| NOIRLab | National Optical-Infrared Astronomy Research Laboratory (NSF center) |
| NPOI | Navy Precision Optical Interferometer |
| NSF | National Science Foundation |
| NUV | Near UltraViolet (wavelength range ~200-400 nm, or alternatively ~300-400 nm) |
| OP | Oxygenic Photosynthesis |
| P3K | PALM3000 adaptive optics system on the Hale 5-m at Palomar Observatory |
| PAG | Program Analysis Group (generic term for ExoPAG, PhysPAG, COPAG) |
| PAL | Present Atmospheric Level |
| PASP | Publications of the Astronomical Society of the Pacific |
| PDR | Preliminary Design Review |
| PEAS | Planet as Exoplanet Analog Spectrograph, an instrument at Lick Observatory |
| PFS | Carnegie Planetary Finder Spectrograph (instrument on Magellan II 6.5-m telescope) |
| PHARO | Palomar High-Angular Resolution Observer (NIR imager used with P3K AO) |
| PhysCOS | Physics of the Cosmos Program |
| PhysPAG | PhysCOS Program Analysis Group (PAG supporting community coordination and analysis for NASA Physics of the Cosmos Program) |
| PI | Principal Investigator |
| PICTURE | Planetary Imaging Concept Testbed Using a Recoverable Experiment |
| PRV | Precision Radial Velocity |
| PSD | Planetary Science Division |
| RASTI | Royal Astronomical Society Techniques & Instruments (journal) |
| RC | Radiative-Convective |
| RCN | Research Coordination Network |
| R&D | Research & Development |
| RFI | Request for Information |
| Robo-AO | Robotic-Adaptive Optics (instrument now on U. Hawai’i 2.2-m telescope) |
| ROSAT | Roengten Satellite (German/US/UK X-ray mission) |
| ROSES | Research Opportunities in Space and Earth Science |
| RV | Radial Velocity |
| SAG | Science Analysis Group |
| SALT | South African Large Telescope |
| SAO | Smithsonian Astrophysical Observatory |
| SAT | Strategic Astrophysics Technology |
| SCExAO | Subaru Coronagraphic Extreme Adaptive Optics |
| SDO | *Solar Dynamics Observatory* |
| SED | Spectral Energy Distribution |
| SETI | Search for Extraterrestrial Intelligence |
| SGL | Science Gap List |
| SIG | Science Interest Group |
| SIT | Science Investigation Team |
| SMARTS | Small & Moderate Aperture Research Telescope System (consortium operating telescopes on Cerro Tololo, Chile, including CTIO/SMARTS 1.5-m telescope) |
| SMD | Science Mission Directorate |
| SMEX | Small Explorer |
| SMP | Single Measurement Precision |
| SOA | State Of the Art |
| SOAR | SOuthern Astrophysical Research (4.1-m telescope at Cerro Pachon, Chile) |
| SOFIA | Stratospheric Observatory for Infrared Astronomy |
| SPA | Science Plan Appendix |
| SPARCS | *Star-Planet Activity Research CubeSat* |
| SPHERE | Spectro-Polarimetric High-contrast Exoplanet REsearch (instrument on VLT) |
| SPHEREx | Spectro-Photometer for the History of the Universe, Epoch of Reionization, and Ices Explorer |
| SPIE | Society of Photo-Optical Instrumentation Engineers |
| START | Science, Technology, Architecture Review Team (2023-2024 group that investigated science cases for HWO; *deprecated*) |
| STDT | Science and Technology Definition Team |
| STIPs | Systems of Tightly-packed Inner Planets (example: TRAPPIST-1) |
| STIS | Space Telescope Imaging Spectrograph (*HST* instrument) |
| STScI | Space Telescope Science Institute |
| TAC | Time Allocation Committee |
| TAG | Technology Assessment Group (2023-2024 group that investigated technologies for HWO; *deprecated*) |
| TBD | To Be Determined |
| TDAMM | Time Domain and Multi-Messenger Astrophysics |
| TESS | *Transiting Exoplanet Survey Satellite* |
| TheoReTS | Theoretical Reims-Tomsk Spectral data (database) |
| TIC | *TESS* Input Catalog |
| TMT | Thirty Meter Telescope |
| TOI | *TESS* Object of Interest |
| TPF | *Terrestrial Planet Finder* (2000s mission concept) |
| TPF-C | *Terrestrial Planet Finder Coronagraph* (2000s mission concept) |
| TPF-I | *Terrestrial Planet Finder Interferometer* (2000s mission concept) |
| TRACE | Technical, Risk, and Cost Evaluation |
| TRAPPIST | Transiting Planets and Planetesimals Small Telescope |
| TSM | Transit Spectroscopy Metric (see Kempton et al., 2018, PASP, 130, 114401) |
| TTV | Transit Timing Variations |
| US-ELTP | US Extremely Large Telescope Program |
| USP | Ultra-Short Period planet |
| UV | UltraViolet |
| VLT | Very Large Telescope |
| VLTI | Very Large Telescope Interferometer |
| VRE | Vegetation Red Edge (surface reflectance biosignature) |
| WASP | Wide Angle Search for Planets |
| WFIRST | *Wide-Field Infrared Survey Telescope* (*deprecated*; previous name for *Nancy Grace Roman Space Telescope* or *Roman Space Telescope* for short) |
| WFS | Wavefront Sensor |
| WG | Working Group |
| WISE | Wide-Field Infrared Survey Explorer |
| WIYN | Wisconsin, Indiana, Yale, NOAO Observatory (host to NEID instrument) |
| WMF | Water Mass Fraction (ratio of mass of water/ice-rich layer to total planet mass) |
| XMM | X-ray Multi Mirror Mission (called “XMM-Newton”; ESA X-ray observatory) |
| XRP | eXoplanet Research Program (an element of the NASA Research Opportunities in Space and Earth Sciences (ROSES) program) |
| XUV | "XUV" in laboratory studies and heliophysics often classically refers to "Extreme Ultraviolet," more commonly acronymed as "EUV" in astronomy, however in recent exoplanet and stellar astronomy literature "XUV" often refers to *combined X-ray and extreme ultraviolet* flux, covering the wavelength range of stellar emission important for photodissociation and ionization in planetary atmospheres. |

# 5. Adopted Exoplanet Terms

A few practical definitions have been adopted for the ExEP Science Gap List which follow guidance from the Astro2020 Decadal Survey and recent influential studies. These are not meant to be exhaustive summaries on these terms, but to provide background on the programmatic use of these terms and point the reader to relevant literature.

## 5.1. "Habitable Zone"

***"Habitable zone"*** **(HZ)** defines *a region around a star where the surfaces of some planets may be able to support liquid surface water*. When otherwise not specified, it corresponds to orbital radii of 0.95-1.67 au for the Sun (Astro 2020 p. 7-16), and for other stars scales as the square root of the bolometric luminosity normalized to the Sun's ($L/L_{Sun}$ ; Kopparapu et al., 2013, ApJ, 765, 131; *LUVOIR* Report p. B-13). Although these limits are called the "***optimistic habitable zone***" in Astro2020, their origin is the "***conservative habitable zone***" from Kasting et al. (1993, Icarus, 101, 108), corresponding to the "*water-loss*" and "*maximum greenhouse*" limits. Kopparapu et al. (2013) derived updated, but similar, limits from newer models, however they recommend that the original 0.95-1.67 au limits "*should be used for current RV surveys and Kepler mission to obtain a lower limit on eta-Earth so that future flagship missions like TPF-C and Darwin are not undersized*." The *LUVOIR* and *HabEx* studies both adopted the 0.95–1.67 au limits and scaled with bolometric luminosity, and this was adopted by Astro2020. The limits of the habitable zone for varying assumptions about the atmospheric composition, water content, rotation, etc. may vary widely, and this definition used here is meant to provide a fiducial definition to guide mission studies.

## 5.2. "Earth-sized"

Astro2020 (p. 2-12, Fig. 7.6, p. 7-16) adopted the exoplanet description ***"Earth-sized"*** to correspond to radii of 0.8-1.4 Earth radii[38]. However the *LUVOIR* and *HabEx* concept studies allowed for smaller planets with lower radius limit of $0.8(a\times(L/L_{Sun})^{0.5})^{-0.5}$ (Kopparapu et al., 2018, ApJ, 856, 122), where $a$ is the orbital semi-major axis in au, $L$ is the stellar bolometric luminosity, $L_{Sun}$ is the IAU nominal solar bolometric luminosity ($3.828\times10^{26}$ W). The lower radius limit corresponds approximately to the "*cosmic shoreline*" where planets above that radius appear to retain their atmospheres, based on an empirical relation between insolation flux and escape velocity (Zahnle & Catling 2017, ApJ, 843, 122). The upper limit of 1.4 $R_{Earth}$ corresponds to the observed limit transition between rocky and gaseous planets (e.g., Rogers 2015, ApJ, 801, 41), however recently this breakpoint has been found to be as high as $1.64\pm0.05$ $R_{Earth}$ (Müller et al., 2024, A&A, 686, A296). The mass-radius relationships for "small" planets (~<4 $M_{Earth}$) are both *predicted* (Zeng et al., 2016, ApJ, 819, 127) and *observed* (Müller et al., 2024, A&A, 686, A296) to follow $R/R_{Earth} \cong (M/M_{Earth})^{0.27}$, suggesting that small planets with the Earth's approximate core mass fraction (CMF ≅ 32%) are common. Planets of Earth-like composition with radii limits of 0.8 $R_{Earth}$ and 1.4 $R_{Earth}$ then correspond to a mass range of approximately 0.4 to 3.5 $M_{Earth}$. There are some rocky planets larger than these upper limits (e.g., 11 $M_{Earth}$ and 1.8 $R_{Earth}$ ; TOI-1347b; Rubenzahl et al., 2024, AJ, 167, 153), and some sub-Neptunes have been found with masses reaching this upper limit (e.g., 4 $M_{Earth}$ and 2.6 $R_{Earth}$; TOI-1266b; Cloutier et al., 2024, MNRAS, 527, 5464). For a star of the Sun's luminosity, the "*cosmic shoreline*" lower radius limit ranges from 0.82 $R_{Earth}$ at the inner HZ edge (0.95 au) to 0.62 $R_{Earth}$ at the outer HZ edge (1.67 au). For the Earth-like composition mass-radius relation, these radii correspond to approximate lower masses of 0.48 $M_{Earth}$ and 0.17 $M_{Earth}$, respectively.

[38] 1 $R_{Earth}$ = "Earth radius" = IAU nominal terrestrial radius = 6378.1 km

## 5.3. "Potentially Habitable Worlds/Planets/Exoplanets"

***"Potentially habitable worlds/planets/exoplanets"*** (Astro2020, pgs. 2-12, 7-16, Fig. 7-6; Exoplanet Science Strategy, pgs. S-2, S-4, 2-3), "***ExoEarth candidates***", (Astro 2020, p. I-2; Kopparapu et al., 2018, ApJ, 856, 122) or "***exo-Earth candidates***" (HabEx report 2019) are then defined to be *"Earth-sized" planets in the "habitable zone"* (following the previous stated constraints in size and orbital semi-major axis). Indeed Astro2020 Fig. 7.6 assumed a 2-parameter definition - "*Habitable zone is defined as 0.95-1.67 AU for planets of 0.8-1.4 Earth radii*" - as their metric for comparing yields as function of telescope diameter for different telescope architectures. The Exoplanet Science Strategy (2018) also generically referred to these planets as "***temperate terrestrial/rocky planets***" (ESS 2018, p. S-2, S-3). The terms are used in the Astro2020 Decadal Survey and Exoplanet Science Strategy (2018) to apply to such planets whether they orbit solar-type stars (e.g., FGK dwarfs) or M dwarfs. These terms are ***not*** meant to imply that a given planet has life or that it is exactly, or even closely, Earth-like in any parameter besides size and instellation or orbital radius (e.g., atmosphere, ocean, tectonics, life, etc.) – *it merely defines a two parameter search space (e.g., planet size and orbital radius) to focus the observational search for signs of habitability (i.e., liquid surface water) and life (i.e., biosignatures) on exoplanets*. The definition has, and will continue to have, *programmatic* importance to NASA ExEP as it is used to define metrics for science yields for comparing mission concept options, in support of the Astro2020 Decadal goal to "*realize a mission to search for biosignatures from a robust number of about ~25 habitable zone planets*" and "*provide a robust sample of ~25 atmospheric spectra of potentially habitable exoplanets*."